\documentclass[aps,prx,twocolumn,longbibliography,superscriptaddress]{revtex4-2}

\usepackage{amsmath, empheq,braket,amsthm,amssymb,graphics,amsfonts,array,color,subfigure, mathrsfs}
\usepackage[unicode=true,pdfusetitle, bookmarks=true,bookmarksnumbered=false,bookmarksopen=false,breaklinks=false,pdfborder={0 0 0},backref=false,colorlinks=true,linkcolor=blue,citecolor=blue,urlcolor=blue]{hyperref}

\usepackage{graphicx,subfigure,lipsum}
\usepackage{graphicx}
\usepackage{dcolumn}
\usepackage{bm}
\usepackage{amsthm} 
\usepackage{hyperref}

\usepackage{lipsum}

\usepackage{algorithm}
\usepackage{algpseudocode}
\usepackage{bm}
\usepackage[dvipsnames]{xcolor}
\usepackage{amssymb}
\usepackage[justification=raggedright, singlelinecheck=false]{caption}
\usepackage{subcaption}
\usepackage{float}

\usepackage{tikz}
\usetikzlibrary{quantikz2}
\usetikzlibrary{graphs}
\usetikzlibrary{fit}

\usepackage{bbm}
\usepackage{physics}
\usepackage{cleveref}
\usepackage[dvipsnames]{xcolor}

\renewcommand{\H}{\mathcal{H}}
\renewcommand{\Tr}[1]{\operatorname{Tr}\!\left[#1\right]}

\newcommand{\create}[1]{\hat{a}_{#1}^\dagger}
\newcommand{\destroy}[1]{\hat{a}_{#1}}
\newcommand{\Fourier}[1][M]{\hat{\mathcal{F}}_{#1}}
\newcommand{\diagcyclic}{\hat{D}_{\hat{S}^{(M)}}}
\newcommand{\diagphase}[1][\phi]{\hat{D}_{#1}}

\preprint{APS/123-QED}

\newcommand{\equalcontrib}{
\thanks{These authors contributed equally to this work.\\
\href{leonardo.finocchiaro@polytechnique.edu}{leonardo.finocchiaro@polytechnique.edu}\\
\href{marco.robbio@ulb.be}{marco.robbio@ulb.be}}
}

\begin{document}

\title{Error Mitigation in Bosonic Systems via Virtual Distillation}

\author{Leonardo Finocchiaro}
\equalcontrib
\affiliation{Department of Microtechnology and Nanoscience, Chalmers University of Technology, G\"{o}teborg SE-412 96, Sweden}
\affiliation{CPHT, LIX, CNRS, Inria, École polytechnique, Institut Polytechnique de Paris, Palaiseau, France}

\author{Marco Robbio}
\equalcontrib
\affiliation{Centre for Quantum Information and Communication, \'Ecole polytechnique de Bruxelles, CP 165/59, Universit\'e libre de Bruxelles, 1050 Brussels, Belgium}
\affiliation{International Iberian Nanotechnology Laboratory (INL), Av. Mestre Jos\'e Veiga, 4715-330 Braga, Portugal}

\author{Diogo Gomes}
\affiliation{International Iberian Nanotechnology Laboratory (INL), Av. Mestre Jos\'e Veiga, 4715-330 Braga, Portugal}
\affiliation{Department of Engineering, University of Minho, R. da Universidade, 4710-057 Braga, Portugal}

\author{David Gunn}
\affiliation{International Iberian Nanotechnology Laboratory (INL), Av. Mestre Jos\'e Veiga, 4715-330 Braga, Portugal}

\author{Adithi Udupa}
\affiliation{Department of Microtechnology and Nanoscience, Chalmers University of Technology, G\"{o}teborg SE-412 96, Sweden}

\author{Axel M Eriksson}
\affiliation{Department of Microtechnology and Nanoscience, Chalmers University of Technology, G\"{o}teborg SE-412 96, Sweden}

\author{Leonardo Novo}
\affiliation{International Iberian Nanotechnology Laboratory (INL), Av. Mestre Jos\'e Veiga, 4715-330 Braga, Portugal}

\author{Giulia Ferrini}
\affiliation{Department of Microtechnology and Nanoscience, Chalmers University of Technology, G\"{o}teborg SE-412 96, Sweden}

\begin{abstract}

Virtual distillation is a promising error-mitigation technique that exploits multiple copies of a noisy quantum state to estimate observables as if measured on a purified state. Although originally introduced in the context of bosonic many-body systems under the name of virtual cooling, its development and applications have largely focused on qubit-based quantum computation. Here, we establish a framework for virtual distillation in bosonic quantum information processing and continuous-variable quantum computing. Building on a diagonalization of cyclic shift operators implemented with passive linear-optical interferometers, we derive experimentally accessible protocols for estimating virtually distilled expectation values of observables relevant to bosonic architectures. In particular, we show how to recover noise-mitigated expectation values of number operators, phase-shift operators, and arbitrary quadratures from multi-copy measurements. For number operators, we further demonstrate the estimation of virtually distilled correlators of arbitrary order through the characteristic function of the photon-number distribution. We apply the framework to states affected by photon loss and dephasing, two of the dominant noise mechanisms in bosonic quantum computation, and quantify the resulting suppression of noise contributions. Our results extend virtual distillation beyond its original setting and provide a practical route toward error-mitigated measurements in bosonic quantum processors using experimentally available linear-optical resources.

\end{abstract}

\maketitle

\section{Introduction}

As quantum computing transitions from the Noisy Intermediate-Scale Quantum (NISQ) era \cite{Preskill2018quantumcomputingin} into the Early Fault-Tolerant Quantum Computing (EFTQC) regime \cite{katabarwa2024early}, mitigating the effects of hardware noise remains a critical bottleneck. The ultimate goal of large-scale quantum error correction promises arbitrarily fault-tolerant quantum computations; however, the hardware overhead entailed by physical redundancy  and encoded logical  operations  is currently prohibitive. In the EFTQC era, near-term architectures will likely feature a limited number of partially corrected logical qubits that still suffer from residual, uncorrected logical errors. 

To maximize the computational utility of these intermediate devices without demanding the full resource scaling of mature fault tolerance, quantum error mitigation \cite{cai2023quantum, zimboras2025mythsquantumcomputationfault}, and in particular, state distillation protocols are an important asset. By extracting higher-fidelity information from multiple noisy copies of a quantum state, these techniques allow us to bridge the gap between raw hardware capabilities and the stringent accuracy demands of reliable quantum algorithms. 

While for qubit systems purity distillation is a well studied field of research \cite{barenco1997stabilization,cirac1999optimal,childs2025streaming,yang2024quantum,keyl2001rate}, the same cannot be said for bosonic quantum computing. So far, purification schemes in this context have been relegated to specific input states, such as Gaussian states \cite{zhang2024purificationgaussianstatesphoton,niset2009no,giedke2002characterization}, or to a specific source of noise in photonic quantum computing known as partial distinguishability \cite{faurbyPurifyingPhotonIndistinguishability2024,hochOptimalDistillationPhotonic2025,marshallDistillationIndistinguishablePhotons2022}. Error mitigation by symmetry expansion has been recently considered for the preparation of the code-vectors of rotationally symmetric bosonic codes \cite{PhysRevA.111.062402}. However, a general framework for bosonic purification protocols is still missing.

A promising framework to address this gap is \emph{Virtual Distillation} (VD) \cite{virtual_cooling,virtual_distillation,koczor2021exponential}, an error-mitigation protocol that uses multiple noisy copies to exponentially suppress noise in estimating observables, as if measured on a purified state. Strikingly, while VD has been successfully developed in qubit-based quantum computation \cite{koczor2021exponential, virtual_distillation}, and even benchmarked and implemented for quantum chemistry tasks \cite{nature_experimental_setup}, the technique was originally born in the bosonic setting \cite{virtual_cooling}.
Its original formulation was motivated by quantum many-body physics and  introduced under the moniker of ``virtual cooling". This  method allows for the estimation of expectation values of a restricted set of observables on a ``virtually cooled" state, by interfering multiple copies of states of higher temperature. Despite originally coming from bosonic systems, the formalism of virtual distillation has not yet been reclaimed for applications in bosonic quantum computing.

In this work, we fill this gap and significantly expand the ``virtual cooling" formalism to allow for virtual distillation of several observables relevant to bosonic quantum computation under realistic noise. We do so  by exploiting a diagonalization technique for cyclic shift operators in linear interferometric networks~\cite{Daley_Diagonalization}. This technique was first introduced to access nonlinear functionals of $\rho$ such as the Rényi entropy, and was successively applied in different contexts 
to probe optical uncertainty, nonclassicality, and entanglement \cite{multicopy-cerf-1,multicopy-cerf-2,multicopy-cerf-3,multicopy-cerf-4}. As a key feature, this diagonalization technique is implementable in terms of passive elements~\cite{multiport_bs, hamiltonian_theory_multiport}, making the virtual distillation protocol feasible with experimentally accessible bosonic circuits. 
While Ref.~\cite{virtual_cooling} focused on estimating the expectation value of the number operator, it failed to directly recover particle number correlations, and needed to introduce a modified correlator whose physical meaning is less understood. In our approach, we show how to obtain noise-mitigated expectation values of quadrature operators and phase-shift operators. The latter allows us not only to reconstruct particle number correlations, but also the full particle number distribution as well as Wigner functions of virtually distilled states. Furthermore, while   Ref.~\cite{virtual_cooling} focused on thermal fluctuations  as the noise source, here we apply  the virtual distillation method to states subject to photon losses and dephasing, which are 
two of the most relevant noise sources in bosonic quantum computation \cite{Joshi_2021}. 
Ultimately, our approach provides a practical route to error-mitigated measurement in bosonic quantum information processing and broadens the scope of virtual distillation to continuous-variable architectures.

The paper is organized as follows: Sec.~\ref{sec: theory} establishes the theoretical formalism of virtual distillation, bosonic systems and virtual cooling. It also introduces   the implementation of VD for bosonic systems  via passive linear interferometric networks using a diagonalization technique for cyclic shift operators. Sec.~\ref{sec:implementation} applies this formalism to the estimation of expectation value of arbitrary quadratures, and introduces a new method for VD based on two linear interference processes. 
Sec.~\ref{sec:results} illustrates, via numerical simulation, the protocols' effectiveness in mitigating photon loss and dephasing for observables like photon number, quadratures, and parity; we also highlight some practical applications, such as Wigner function reconstruction and enhanced parity measurements. 
In the Appendices, we provide the proof of the presented analytical results, derive a full example of interferometer for virtual distillation, and assess the robustness of the protocol to additional noise that may occur in a physical setup during the distillation process itself.

\section{Theoretical background}\label{sec: theory}

In this Section, we briefly introduce virtual distillation, our notation for bosonic systems, and summarize the prior work on virtual cooling, also adapting it to the context of quantum information with bosonic systems. 

\subsection{Virtual distillation}\label{sec:background_VD}

Consider an observable \( \hat{O} \) whose expectation value we wish to evaluate on the pure target state \( \ket{\psi_{0}} \). Suppose that we only have access to copies of a noisy state \( \rho \), which is assumed to be close to \( \ket{\psi_{0}} \) and can be expressed as
\begin{equation}\label{eq:noisy-input}
\rho = (1-\epsilon)\ketbra{\psi_{0}} + \sum_i \epsilon_{i} \ketbra{\psi_i} \quad \text{s.t. } \langle \psi_{i}|\psi_{j}\rangle=\delta_{ij},
\end{equation}
where \( 1-\epsilon > \epsilon_{i}  ,\ \forall i\) (i.e., \( \ket{\psi_{0}} \) is the dominant eigenvector). A purified version of $\rho$ is given by
\begin{equation}
\tilde{\rho} := \frac{\rho^M}{\Tr{\rho^M}} = \frac{(1-\epsilon)^M \ketbra{\psi_{0}} + \sum_i \epsilon_{i}^M \ketbra{\psi_i}}{(1-\epsilon)^M + \sum_i \epsilon_{i}^M},
\end{equation}  
where $\rho^M$ represents the $M$-th power of $\rho$.
The dominant eigenvector of $\tilde\rho$ remains \( \ket{\psi_{0}} \), but its weight is enhanced. Thus, expectation values computed with respect to $\tilde{\rho}$, in the limit of large $M$, provide a better approximation to $\bra{\psi_{0}}\hat{O}\ket{\psi_{0}}$ than directly with \(\rho\). Indeed, in the limit of large \( M \),
\begin{equation}\label{eq: main idea}
\Tr{\hat{O}\tilde{\rho}}=\frac{\Tr{\hat{O} \rho^M}}{\Tr{\rho^M}} \approx \bra{\psi_{0}}\hat{O}\ket{\psi_{0}}.
\end{equation}
The error suppression with respect to the parameter $M$ is exponential; for $\frac{1}{2}\bigl\|\rho - |\psi_{0}\rangle\langle \psi_{0}|\bigr\|_{1} = \mathcal{O}(\epsilon)$, we have
\begin{equation}
\label{eq:operator-norm-bound}
    \frac{1}{2}\left|\Tr{\hat{O}\left(\tilde{\rho} - |\psi_{0}\rangle\langle \psi_{0}|\right)}\right|
    \leq \|\hat{O}\|_{\infty}\,\mathcal{O}(\epsilon^{M}),
\end{equation}
by Hölder's inequality \cite{rudinRealComplexAnalysis2013}. 
Nevertheless, preparing the state $\tilde{\rho}$ from $\rho$ is not easy and requires in the qubit model auxiliary qubits and post-selection, see Refs.~\cite{barenco1997stabilization,yang2024quantum}. 

In a virtual distillation (VD) protocol \cite{virtual_cooling,virtual_distillation}, one does not prepare $\tilde{\rho}$. 
Instead, VD directly estimates the quantities \( \Tr{\hat{O} \rho^M} \) and \( \Tr{\rho^M} \) in Eq.~\eqref{eq: main idea} with measurements on $M$ copies of $\rho$. In Refs.~\cite{virtual_cooling, virtual_distillation} two approaches are developed to estimate $\Tr{\hat{O} \rho^M}$ based on the two following identities:
\begin{align}
    \Tr{\hat{O}\rho^M}&=\Tr{\hat{O}^{(M)}\hat{S}^{(M)}\rho^{\otimes M}}  \label{eq:Approach1}\\
     \Tr{\hat{O}\rho^M}&=\Tr{\hat{O}_{k}\hat{S}^{(M)}\rho^{\otimes M}}, \label{eq:Approach2}
\end{align}
where $\hat{S}^{(M)}$ is the cyclic shift operator acting on $\H^{\otimes M}$ as
\begin{equation}
\hat{S}^{(M)} |\psi_{1},\psi_{2},\dots,\psi_{M}\rangle=|\psi_{2},\dots,\psi_{M},\psi_{1}\rangle \ ,
\end{equation}
$\hat{O}_{k}=\mathbb{I}^{\otimes (k-1)} \otimes \hat{O} \otimes \mathbb{I}^{\otimes (M-k)}$ and $\hat O^{(M)}$ is the symmetrized version of $\hat{O}$ over all the modes,
\begin{equation}
    \hat{O}^{(M)}=\frac{1}{M}\sum_{k=1}^{M} \hat{O}_{k}.
\end{equation}

To evaluate the denominator of Eq.~\eqref{eq: main idea}, one can simply plug $\hat O=\mathbb I$ into Eq.~\eqref{eq:Approach1}. Thus, estimating \(\langle \psi_0|\hat{O}|\psi_0\rangle\) amounts to measuring the expectation values of \(\hat{S}^{(M)}\), and either \(\hat{O}^{(M)}\hat{S}^{(M)}\) or \(\hat{O}_{1}\hat{S}^{(M)}\) on the state $\rho^{\otimes M}$. Since both quantities yield the same result, the choice of approach depends on which operator is easier to implement in an actual setup. In Sec.~\ref{sec:implementation}, we will see how different scenarios are better suited for one approach or the other.
Note that the estimator of $\langle\psi_{0}|\hat{O}|\psi_{0}\rangle$ in Eq.~\eqref{eq: main idea} is the ratio of two quantities that must, in general, be estimated separately. 
Both quantities decrease exponentially in the number of copies. Consequently, the number of samples required for a precise estimation scales exponentially in $M$ (for a detailed discussion, see Appendix~\ref{sec: Sample complexity}).

One should note that the effectiveness of virtual distillation as an error mitigation protocol depends on the validity of the assumption on the effect of the noise in Eq.~\eqref{eq:noisy-input}.
This raises the question of whether this condition is fulfilled under standard noise models. It turns out that the answer strongly depends on the state and noise type.
In general, the noise may modify the dominant eigenvector, resulting in an input state $\rho$ whose eigendecomposition is not of the form of Eq.~\eqref{eq:noisy-input}. In this case, the protocol will compute expectation values on the dominant eigenvector of $\rho$, which may be different from the target pure state $\ket{\psi_0}$.
We refer to this issue as \emph{eigenvector drift}, as the dominant eigenvector of $\rho$ is drifting away from the target $\ket{\psi_0}$ as a function of noise strength. This scenario has been extensively studied in the context of qubit virtual distillation in Ref.~\cite{Koczor_2021}. When $\ket{\psi_0}$ is an eigenstate of the target observable $\hat O$, the effect of the eigenvector drift on the result of the protocol is exponentially less severe with respect to the incoherent loss $\epsilon$. We will show  that this effect is also  present for VD of bosonic systems.

\subsection{Bosonic systems}

Virtual distillation can be applied in qubit \cite{virtual_distillation} or bosonic settings. 
In this work, we consider the latter.
Let us therefore introduce the relevant bosonic notation.
An $M$-mode \emph{bosonic system}, or \emph{continuous-variable (CV) system} is described by the tensor product of $M$ single-mode Fock spaces, each of them associated with creation/annihilation operators $\create{i},\destroy{i}$, where $i\in\{1,...,M\}$ indexes the mode. 
The operators satisfy the bosonic canonical commutation relations
\begin{equation}
    [\destroy{i},\create{j}]=\delta_{ij} \quad ,\quad [\destroy{i},\destroy{j}]=0,
\end{equation}
which are used to define the number operator of the $i$-th mode  $\hat{n}_i=\create{i} \destroy{i}$.
For each mode $j$, the  quadrature operators $\hat q_j$ and $\hat p_j$ are defined as
\begin{align}
\hat q_j = \frac{\hat a_j + \hat a_j^\dagger}{\sqrt{2}} \quad , \quad \hat p_j = \frac{\hat a_j - \hat a_j^\dagger}{i\sqrt{2}}.
\end{align}
Some usual states considered in bosonic protocols include Fock states $\ket n$ for $n\in\mathbb N$, which are the eigenvectors of the number operator $\hat n$, and coherent states $\ket\alpha$ for $\alpha\in\mathbb C$ which are eigenvectors of the annihilation operator $\hat a$. A multi-mode system can correspond to multiple physical systems (such as optical or microwave cavities), or to a single system with several degrees of freedom, yielding bosonic multiple modes. Both scenarios will be relevant in the following, especially in Sec.~\ref{sec:multi-mode}. 

Passive linear interferometers acting on $M $ modes will play a central role in our protocols. 
These unitaries are characterized by their action on the annihilation operators:

\begin{equation}
    \hat{U}^{\dagger}\destroy{i}\hat{U}=\sum_{j}U_{i,j}\destroy{j},
\end{equation}
where $\hat U$ is an $M$-dimensional unitary matrix. Such unitary transformations are passive, meaning that they conserve the total number of particles, and can always be decomposed into $O(M^{2})$ Mach-Zehnder interferometers with limited connectivity \cite{reckExperimentalRealizationAny1994,clementsOptimalDesignUniversal2017}. 
In the following, we will make heavy use of the fact that any linear interferometer  $\hat{U}$ can be diagonalized as  
    \begin{equation}\label{prop: Diagonalization}
        \hat{U}=\hat{V}_U^{\dagger}\hat{D}_U\hat{V}_U,
    \end{equation}
    where $\hat{V}_U$ is a linear interferometer, and $\hat{D}_U$ is an operator diagonal in the Fock basis, which can be written as
    \begin{equation}
\hat{D}_U=e^{i\sum_{j}\phi_{j}\hat{n}_{j}}=\prod_{j}e^{i\phi_{j}\hat{n}_{j}},
    \end{equation}
    for some set of phases $\{\phi_j\}$.
Finding the operators $\hat{V}_U$ and $\hat{D}_U$ is not computationally expensive, since it reduces to diagonalizing the associated $M \times M$ unitary matrix $U$, which can be done in $O(M^{3})$ operations on a classical computer.

\subsection{Virtual cooling and continuous variable virtual distillation}

Having introduced bosonic systems, we summarize here the idea of the virtual cooling protocol from Ref.~\cite{virtual_cooling}, which can be  directly applied to continuous variable virtual distillation. There, they use $M$ copies of a thermal state at temperature $T$ to estimate expectation values at the lower effective temperature $T/M$.
Specifically, for a state in thermal equilibrium, we have
\begin{equation}
    \rho_{th}=\frac{e^{-\beta H}}{\Tr{e^{-\beta H}}} \implies \tilde{\rho}_{th}=\frac{e^{-M\beta H}}{\Tr{e^{-M\beta H}}}
\end{equation}
which is equivalent to the transformation $\beta\to M\beta$, thereby simulating a temperature reduction of a factor of $M$. 
Here we review the protocol of Ref.~\cite{virtual_cooling} in the context of continuous variable virtual distillation.

The central idea is that $\hat{S}^{(M)}$ can be diagonalized as

\begin{equation}
    \hat{S}^{(M)}=\hat{\mathcal F}_M^\dagger\,\diagcyclic\,\hat{\mathcal F}_M,
\end{equation}
where $\Fourier$ is the $M$-discrete Fourier transform interferometer
\begin{equation}\label{eq:def_Fourier}
    \hat{\mathcal F}_M^\dagger \destroy{j} \hat{\mathcal F}_M = \frac{1}{\sqrt{M}} \sum_{k=1}^M e^{ \frac{2i\pi k}{M}j}\destroy{k} 
\end{equation}
and $\diagcyclic$ is a layer of phase shifters,
\begin{align}\label{eq:diagonal shift}
     \diagcyclic &=\exp\!\left(  \frac{2i\pi}{M} \sum_{j=1}^{M} j\,\hat n_j \right)   .
\end{align}
As a consequence, $\Tr{\hat{O}\rho^{M}}$ can be estimated as follows:
\begin{align}
    \Tr{\hat{O}\rho^{M}}&=\Tr{\hat{O}^{(M)}\hat{S}^{(M)}\rho^{\otimes M}}\nonumber\\
    &=\Tr{\hat{O}^{(M)}\hat{\mathcal F}_M^\dagger\,\diagcyclic\,\hat{\mathcal F}_M\rho^{\otimes M}}\nonumber\\
    &=\Tr{\hat{\mathcal F}_M\hat{O}^{(M)}\hat{\mathcal F}_M^\dagger\,\diagcyclic\,\hat{\mathcal F}_M\rho^{\otimes M}\hat{\mathcal F}_M^{\dagger}}\nonumber\\
    &=\langle \hat{\mathcal F}_M\hat{O}^{(M)}\hat{\mathcal F}_M^\dagger\,\diagcyclic \rangle_{\hat{\mathcal F}_M\rho^{\otimes M}\hat{\mathcal F}_M^{\dagger}}.\label{eq:numerator_Fourier}
\end{align}

The denominator of Eq.~\eqref{eq: main idea} is obtained in a similar way by setting $\hat O=\mathbb{I}$, in which case Eq.~\eqref{eq:numerator_Fourier} reduces to $\langle \diagcyclic \rangle_{\hat{\mathcal F}_M\rho^{\otimes M}\hat{\mathcal F}_M^{\dagger}}$. As
$\diagcyclic$ is diagonal in the Fock basis, 
this expectation value can be recovered by evolving $\rho^{\otimes M}$ through a Fourier interferometer and measuring the particle number in all modes.
If we can also estimate $\hat{\mathcal F}_M\hat{O}^{(M)}\hat{\mathcal F}_M^\dagger\,\diagcyclic $ with an efficient scheme, then we can obtain the expectation value we are aiming at. 

In Ref.~\cite{virtual_cooling}, particular attention is devoted to the case of computing the expectation value of the particle number operator $\hat n$. The key observation that allows this is that $\hat{n}^{(M)}=\frac{1}{M}\sum_{i=1}^{M}\hat{n}_{i}=\frac{\hat{N}_{tot}}{M}$, and, since a passive linear unitary does not alter the total number of particles, we have that $ \hat{\mathcal F}_M\hat{n}^{(M)}\hat{\mathcal F}_M^\dagger=\hat{n}^{(M)}$. Inserting this in Eq.~\eqref{eq:numerator_Fourier} yields 
\begin{equation} \label{eq: diagonalizing number op}
    \Tr{\hat{n}\rho^{M}}
    =\langle \hat{n}^{(M)} \diagcyclic \rangle_{\Fourier \rho^{\otimes M} \Fourier^\dagger}.
\end{equation}
Notice that both $\hat{n}^{(M)}$ and $\diagcyclic$ are diagonal in the Fock basis.
Therefore, the task simply reduces to sampling in the Fock basis and then classically post-processing the data to obtain the expectation value, see Appendix~\ref{sec: example} for a detailed example.
Note that the sampling task gives the data for both observables $\hat{S}^{(M)}$ and $\hat{n}^{(M)}\hat{S}^{(M)}$ simultaneously; calculating the two expectation values only requires different classical post-processing. Thus, this approach only requires a single interferometer.

\section{Protocols for Virtual Distillation with bosonic systems}\label{sec:implementation}

\begin{figure*}[!htbp]
\centering
  \begin{minipage}[c]{0.45\textwidth}
  \centering
    \includegraphics[width=\textwidth]{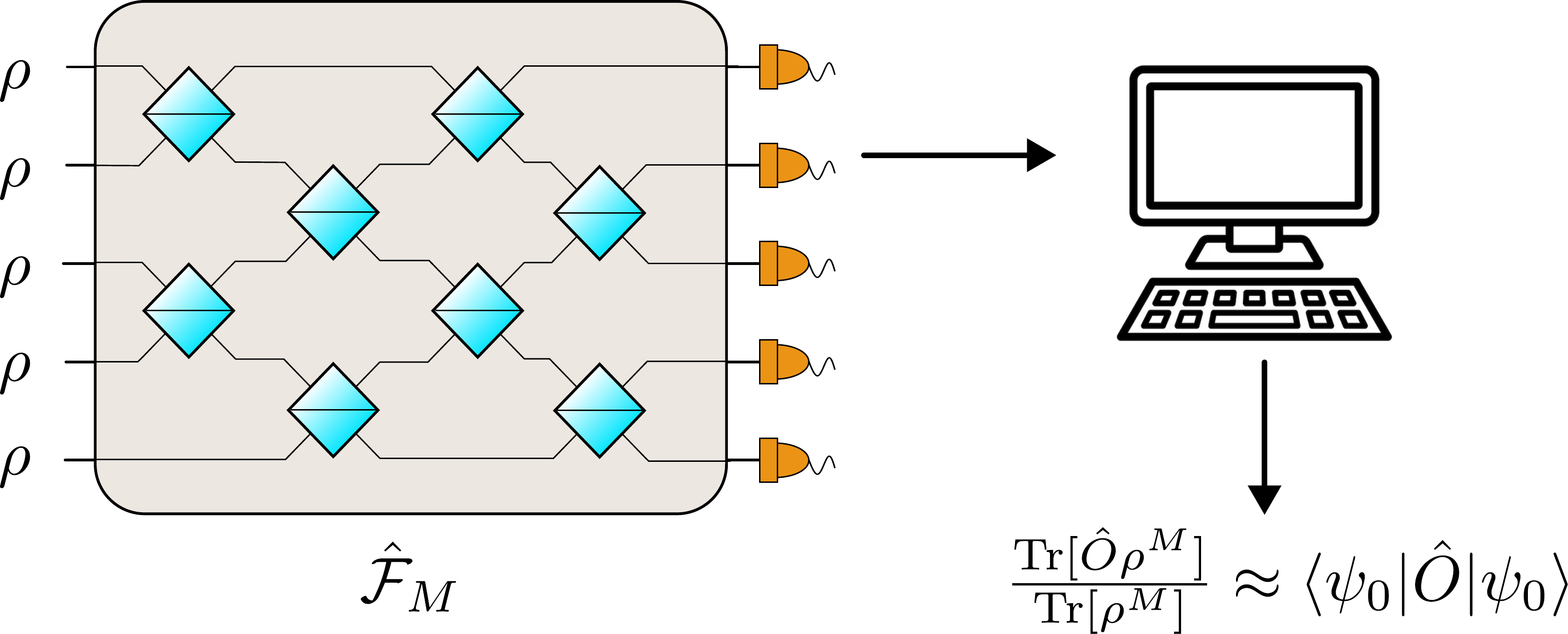}
    \caption*{(a)}
  \end{minipage}
  \hfill
  \begin{minipage}[c]{0.45\textwidth}
  \centering
    \includegraphics[width=\textwidth]{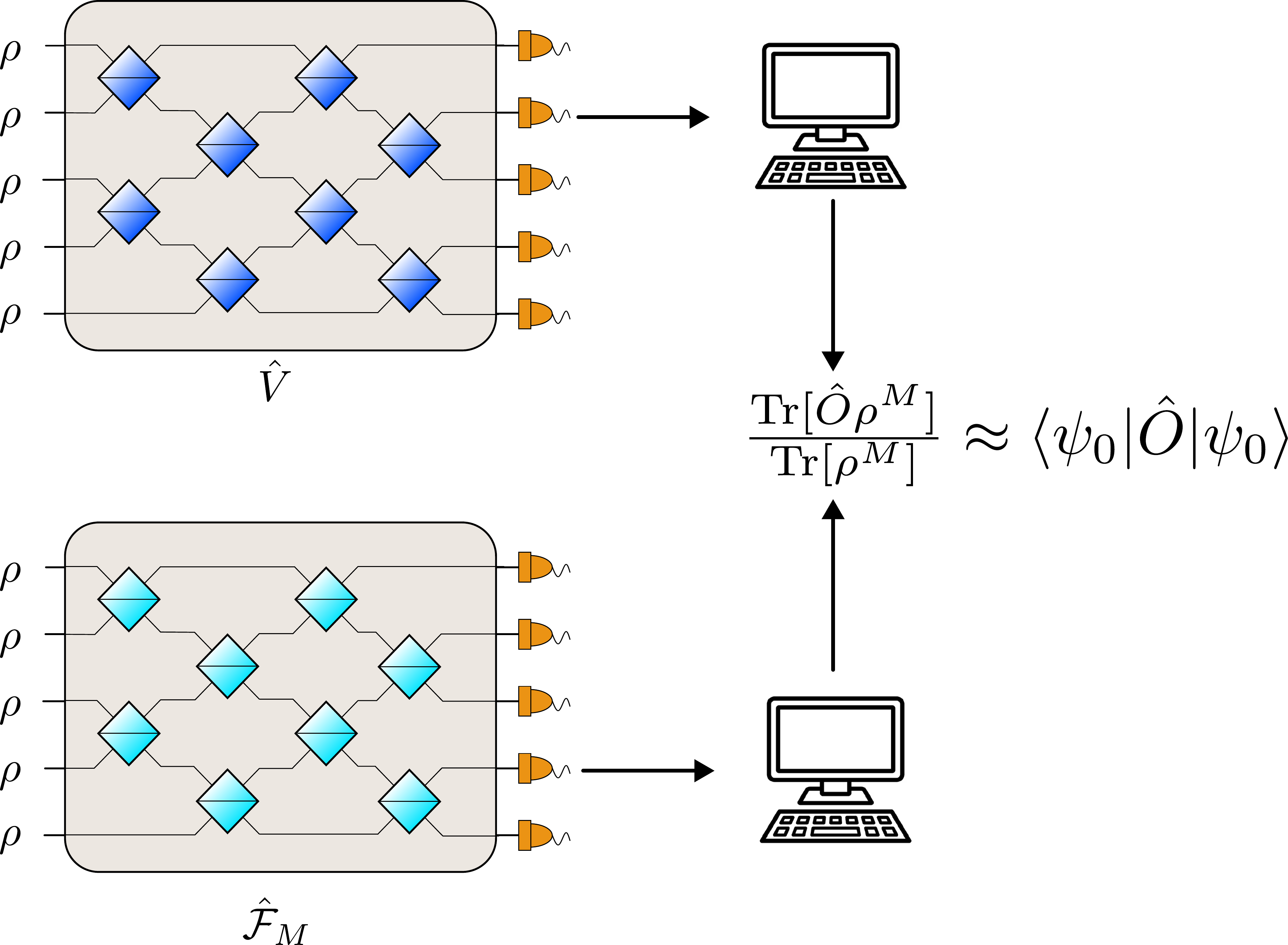}
    \caption*{(b)}
  \end{minipage}
  \caption{Schematic of the two possible approaches to virtual distillation using the diagonalizing technique: (a) The $M$ copies of the noisy state $\rho$ are combined through a  Fourier interferometer $\hat{\mathcal F}_M$, enabling the simultaneous estimation of $\Tr{\hat{O}\rho^M}$ and $\Tr{\rho^M}$ from joint measurements on the output modes; (b) Two interferometers are needed to perform the protocol:  the Fourier interferometer $\hat{\mathcal F}_M$ for the estimation of $\Tr{\rho^M}$, and another interferometer $\hat V$ depending on the observable $\hat{O}$ for the estimation of $\Tr{\hat{O}\rho^M}$. 
  In general, any linear unitary acting on $M$ modes can be decomposed in terms of $O(M^{2})$ passive linear elements, i.e. beam-splitters and phase shifters, with limited connectivity, meaning only coupling between modes $(i,i+1) \ \forall i$  \cite{clementsOptimalDesignUniversal2017,reckExperimentalRealizationAny1994}.  In the case of the Fourier interferometer, this number of elements can be reduced to $O(M\ln M)$ if it is possible to access arbitrary coupling \cite{marcandelli2026continuous}.}
  \label{fig:Protocols}
\end{figure*}

In this Section, we develop two methods, depicted in  Fig.~\ref{fig:Protocols}, for mitigating errors when estimating the expectation value of bosonic observables, using Eq.~\eqref{eq:Approach1} and Eq.~\eqref{eq:Approach2} respectively. 
Method~1 is inspired by those outlined in Refs.~\cite{virtual_cooling, virtual_distillation} and requires only a single Fourier interferometer and mode-local measurements. 
Method~2 uses the fact that any linear interferometer can be diagonalized (see Eq.~\eqref{prop: Diagonalization}) and requires two interferometers and particle-number measurements on each mode.

Before presenting the details of the protocol, let us start by discussing the bounds on its efficiency. In CV systems, many physically relevant observables are unbounded, such as the quadrature operators $\hat{q}$, $\hat{p}$, and the number operator $\hat{n}$. 
Therefore, the operator norm in  Eq.~\eqref{eq:operator-norm-bound} is often unsatisfactory in this framework, and we need to derive an alternative bound, as compared to the one in Eq.~\eqref{eq:operator-norm-bound}.
By using Eq.~\eqref{eq:noisy-input}, the error can be bounded as follows:
\begin{equation}\label{eq: bound via maxi}
    \left|\Tr{\hat{O}\left(\tilde{\rho} - |\psi_{0}\rangle\langle \psi_{0}|\right)}\right|
    \leq 2\max_{i}\left|\Tr{\hat{O}|\psi_{i}\rangle\langle\psi_{i}|}\right|\,\mathcal{O}(\epsilon^{M})
\end{equation}
where the coefficient $\Tr{\hat{O}|\psi_{i}\rangle\langle\psi_{i}|}$ can be assumed to be bounded; see Appendix~\ref{sec: Bound} for a detailed derivation. We now turn to the details of the two methods of virtual distillation that we consider. 

\subsection{Method~1: Virtual Distillation using one interferometer}

The original idea developed in \cite{virtual_cooling, virtual_distillation} makes use of a single Fourier interferometer to compute the expectation value of specific observables such as the number operator.
Here, we extend this approach to compute expectation values of any quadrature operator. We will refer to this first approach as Method~1, which is visualized in Fig.~\ref{fig:Protocols}(a).

To this end, consider as an example the observable $\hat q$. 
The key observation is that
\begin{align*}
    \Fourier \hat q^{(M)}\Fourier^\dagger&=\frac1{\sqrt M}\hat q_M,
\end{align*}
where $\hat q_M$ is the quadrature operator acting on mode $M$ only.
Therefore, 
\begin{equation}
\Tr{\hat{q}\rho^M}=\left\langle \frac{\hat{q}_M}{\sqrt{M}}\diagcyclic \right\rangle_{\Fourier \rho^{\otimes M} \Fourier^\dagger}.
\label{eq:QuadratureExpValue}
\end{equation}

To measure the right-hand side of Eq.~\eqref{eq:QuadratureExpValue}, we can perform a particle number measurement on the first $M-1$ modes and a homodyne measurement on mode $M$, since the operator $\diagcyclic$ only involves $\hat n_1,\dots, \hat n_{M-1}$ (see note \footnote{By Eq.~\eqref{eq:diagonal shift}, we see that the operator $\diagcyclic$ only involves the $M-1$ first modes, since the last one is associated with the observable $e^{\frac{2i\pi}{M}M\hat{n}_{M}}=1^{\hat{n}_{M}}$ which does not affect the result.}).
An analogous expression holds for $\hat p$, where the final measurement on the $M$-th mode is replaced by $\hat p_M$. Since the expressions for $\langle \hat q^{(M)} \hat{S}^{(M)} \rangle$ and $\langle \hat p^{(M)} \hat{S}^{(M)} \rangle$ are linear in $\hat q$ and $\hat p$, we have access within this virtual distillation protocol to arbitrary quadrature operators defined as
\begin{equation}\label{eq:arbitrary-quadrature}
    \hat q_\theta = \cos\theta \,\hat q + \sin\theta\,\hat p.
\end{equation}
The denominator is treated as in the previous setup, by post-processing the value obtained by measurement of the number operator on the first $M-1$ modes.

As noted previously, this first method requires only one circuit implementing the Fourier interferometer to prepare the state used to extract the desired expectation values. 
Moreover, for the observables considered, measurements are then performed on each mode individually. 
The expectation value on the distilled state is inferred by classically post-processing the measurement outcomes. 
As expected, the distilled state is never actually prepared; rather, its behavior is mimicked by suitable measurements on $\hat{\mathcal F}_M\rho^{\otimes M}\hat{\mathcal F}_M^\dagger$.

\subsection{Method~2: Virtual Distillation using two interferometers}\label{sec:Method2}

While Method~1 works well for the observables $\hat n$, $\hat q$, as well as arbitrary quadratures, the operator $\hat{\mathcal F}_M \hat O^{(M)} \hat{\mathcal F}_M \diagcyclic$ may be impractical to measure for a general observable $\hat O$.
For example, for the case of the parity operator $(-1)^{\hat{n}}$, we obtain
\begin{equation}
    \hat{\mathcal F}_M \hat O^{(M)} \hat{\mathcal F}_M^\dagger
= \frac1M\sum_{i=1}^M(-1)^{\frac1M\sum_{k,l=1}^M\omega^{(l-k)i}\hat a_k^\dagger\hat a_l} \ ,
\end{equation}
with $\omega=\exp(2 i\pi/M)$. 
This is a highly non-local operator, not diagonal in the Fock basis, and it is therefore  impractical to measure. 

Since Method~1 is not well suited for accessing expectation values of general observables such as parity, we introduce an alternative interferometric scheme shown in Fig.~\ref{fig:Protocols}(b).
This scheme can be used to estimate expectation values of arbitrary passive linear unitary operators. 
As a particularly relevant example, we focus on the continuous family of phase shift operators $e^{i\phi \hat{n}}$.
Note that this family includes the parity operator $(-1)^{\hat{n}}$.
We notice that $e^{i\phi\hat{n}_{1}}\hat{S}^{(M)}$ is a linear operator and thus can be diagonalized according to Eq.~\eqref{prop: Diagonalization}. Therefore, we can write
\begin{equation}\label{eq:phase shifter diagonalization}
    e^{i\phi\hat{n}_{1}}\hat{S}^{(M)}=\hat{V}_{\phi}^{\dagger}\diagphase\hat{V}_{\phi},
\end{equation}
where we used the index $\phi$ to recall the dependence on the phase $\phi$ of the phase shifter. As a consequence, we can rewrite 
\begin{align}\label{eq: phase formula}
    \frac{\Tr{\hat{O}\rho^{M}}}{\Tr{\rho^{M}}}&=\frac{\Tr{\diagphase\hat{V}_{\phi}\rho^{\otimes M}\hat{V}_{\phi}^{\dagger}}}{\Tr{\diagcyclic\Fourier \rho^{\otimes M}\Fourier^\dagger}}\\
    &=\frac{\langle \diagphase\rangle_{\hat{V}_{\phi}\rho^{\otimes M}\hat{V}_{\phi}^{\dagger}}}{\langle \diagcyclic\rangle_{\hat{\mathcal{F}}_{M}\rho^{\otimes M}\hat{\mathcal{F}}_{M}^{\dagger}}}.
\end{align}
As $\diagphase$ and $\diagcyclic$ are diagonal in the Fock basis, it is clear that collecting samples from measurements in the Fock basis after evolving $\rho^{\otimes M}$ through the respective interferometers $\hat{V}_\phi$ and $\Fourier$, one can estimate the desired expectation value. We refer to this as Method~2.

We emphasize that, in this case, the data from one interferometer is not sufficient to evaluate the numerator and the denominator.
Instead, the two must be evaluated separately with two different experiments.
Nevertheless, this additional interferometer can also be decomposed into $O(M^{2})$ Mach-Zehnder interferometers with limited connectivity, and therefore the additional experimental cost is limited. In Appendix~\ref{sec: Analytical form}, we provide an analytical form for $\hat{V}_{\phi}$ and $\hat{D}_{\phi}$.
Note also that the data used to estimate the denominator in Eq.~\eqref{eq: phase formula} can also be used to estimate some observables via Method~1.

We also notice that, by  having access to the expectation value of the phase shift operator $e^{i \phi \hat n}$, we can also recover the characteristic function (Fourier transform) of the photon number probability distribution, $\{p_\rho(n)\}_{n\in\mathbb{N}}$, defined as
\begin{align}
    \chi(\phi)&\coloneq\sum_n p_\rho(n)e^{i\phi n}=\sum_{n}\Tr{|n\rangle\langle n|\rho}e^{i \phi n}\\
    &=\Tr{e^{i \phi \hat{n}}\rho}.
\end{align}
If the state $\rho$ has a bounded support probability distribution, i.e. $\exists \ N : \Tr{|k\rangle\langle k|\rho}=0 \ \forall k\geq  N$, then it is possible to invert the photon-number characteristic function to get access to the probability distribution
\begin{equation}
  p_{\rho}(n)=\sum_{l=0}^{N-1}\chi\left(\frac{2\pi}{N}l\right)e^{-\frac{2i\pi}{N}l\cdot n} \ .
\end{equation}
Therefore, estimating the photon-number characteristic function in $N$ points allows for accessing the entire photon number distribution by inversion with the Discrete Fourier transform. Alternatively, for any finite-energy state, it is in general possible to find a truncation that guarantees an exponentially small  error \cite{upreti2026exponentiallyimprovedeffectivedescriptionsphysical,bressanini2024gaussianbosonsamplingvalidation}. 
The photon-number characteristic function has  multiple applications in the context of bosonic systems, where it has been used both for analytical calculations \cite{robbio2026complementaritybosonicfermionicmanybody,van2023majorization} and as a technique to compute binned distribution used as validation tool for boson sampling experiments \cite{anguita2025experimentalvalidationbosonsampling,bezerra2026generating,bressanini2024gaussianbosonsamplingvalidation}.

\subsection{Multi-mode generalization}\label{sec:multi-mode}

Until now, we have considered the case where $\ket{\psi_0}$ is a state in a single-mode Fock space, and considered schemes involving $M$ copies of this single-mode state. 
To generalize this scheme to multimode states, we adopt the same approach as in Ref.~\cite{novoNativeLinearopticalProtocol2026}. When several copies and several modes are explicitly considered at the same time, we adopt the notation for the creation operator $a_{i,\alpha}^{\dagger}$, whose two indices indicate, respectively, the copy label (Latin index $i$) and the mode label (Greek index $\alpha$). We will refer to the former as external degrees of freedom, labeling which of the $M$ copies of $\rho$ we are considering, and to the latter as internal degrees of freedom, labeling which mode of that copy of $\rho$ we are considering. We say that a linear interferometer does not act on the internal degrees of freedom if its action can be described as
\begin{equation}
    \hat{U}^{\dagger}\destroy{i,\alpha}\hat{U}=\sum_{j}U_{i,j}\destroy{j,\alpha}.   
\end{equation}
In particular, we can build a linear interferometer implementing the cyclic shift between multimode states as
\begin{equation}    
(\hat{S}_{\operatorname{mult}}^{(M)} )^{\dagger}\destroy{i,\alpha}\hat{S}_{\operatorname{mult}}^{(M)}=\destroy{i+1 (\operatorname{mod} M),\alpha},
\end{equation}
where we use the index ``$\operatorname{mult}$" to indicate the multimode nature of the cyclic shift. One can show that the matrix that diagonalizes this interferometer is a layer of Fourier interferometers acting on the same internal degrees of freedom $\alpha$ (a detailed description is given in Ref.~\cite{novoNativeLinearopticalProtocol2026}), and the associated diagonal operator can be written as
\begin{equation}
    \hat D_{S_{\operatorname{mult}}^{(M)}}=\prod_{j=1}^{M}e^{\frac{2i\pi}{M}j\sum_{\alpha}\hat{n}_{j,\alpha}}   .
\end{equation}
Therefore,  the same schemes described above can also be used in  the multimode case. In particular, if we are interested in the expectation value of the observable $\hat{n}_{\alpha}$, we can express it as 
\begin{equation}
    \Tr{\hat{n}_{\alpha}\rho^{M}}=\frac{1}{M}\langle \left(\sum_{j=1}^{M} \hat{n}_{j,\alpha}\right)\hat{D}_{\hat{S}^{(M)}_{\operatorname{mult}}} \rangle_{\hat{\mathcal F}_{M,\operatorname{mult}}\rho^{\otimes M}\hat{\mathcal F}_{M,\operatorname{mult}}^{\dagger}} \ .
\end{equation}
The same can be said for the operators $\hat{q}_{\alpha},\hat{p}_{\alpha},e^{i\phi\hat{n}_{\alpha}}$. 

Notably, the estimation of expectation values of products of phase-shift operators $e^{i\phi\hat{n}_{\alpha}}$ enables the estimation of particle number distributions over subsystems (marginal or binned distributions) \cite{seron2024efficient}, as well as particle number correlators—such as $\langle\hat{n}_{\alpha}\hat{n}_{\beta}\rangle$ and higher-order terms. These can be obtained through the properties of the photon-number characteristic function, whose multimode generalization is defined as 
\begin{align}
\chi(\phi_{\alpha_{1}},...,\phi_{\alpha_{k}})=\Tr{e^{i \sum_{j}\phi_{\alpha_{j}} \hat{n}_{\alpha_{j}}}\rho} .
\end{align}
Generalizing Eq.~\eqref{eq:phase shifter diagonalization}, we can then adapt Method~2 to the multimode case to estimate the characteristic function at several points and deduce from it the desired correlator.
This capability represents a significant advantage over the original virtual cooling proposal \cite{virtual_cooling}, where estimating two-mode correlators without an auxiliary qubit was not directly feasible.
Indeed, attempting to extract these correlators directly, using only the Fourier transform approach, fails to isolate the desired terms, yielding for two copies
\begin{align}
    \Tr{ \left(\hat{n}_{\alpha}\hat{n}_{\beta}\right)^{(2)} \hat{S}^{(2)}_{\operatorname{mult}}\rho^{\otimes 2}}&=\frac{1}{2}\Tr{\hat{n}_{\alpha}\hat{n}_{\beta}\rho^{2}}+\frac{1}{2}\Tr{\hat{n}_{\alpha}\rho\hat{n}_{\beta}\rho} \nonumber\\
    &\neq \Tr{\hat{n}_{\alpha}\hat{n}_{\beta}\rho^{2}}   .
\end{align}

By contrast, the characteristic function circumvents this issue and allows us to estimate arbitrary noise-mitigated correlators cleanly. In Appendix~\ref{sec: Moment estimation}, we discuss the available approaches for obtaining these correlators—analyzing the required number of measurement setups and sampling errors—and numerically demonstrate the efficiency of this specific task via an example inspired by recent atomic boson sampling experiments~\cite{Young_2024} . Practically speaking, access to arbitrary correlators provides a direct pathway for near-term quantum devices to observe complex many-body physics phenomena, such as long-range correlations and phase transitions, that would otherwise be obscured by hardware noise.

\section{Numerical simulations and applications}\label{sec:results}

\begin{figure*}[!htbp]
    \centering
    \includegraphics[width=\linewidth]{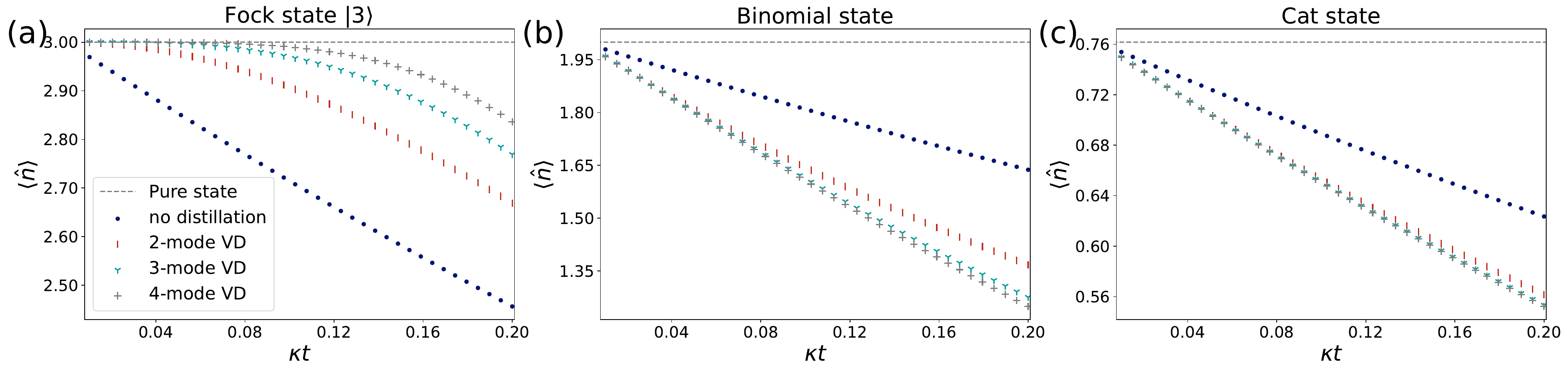} 
    \caption{Expectation value of the number operator as a function of the dimensionless loss rate, $\kappa t$, for inputs (a) Fock state $|3\rangle$; (b) binomial state $|\psi_{b}\rangle\propto|0\rangle+|4\rangle$ and (c) cat state $|\psi_{c}\rangle\propto|\alpha\rangle+|-\alpha\rangle$. Compared to the reference plot without distillation, the protocol performs at its best on the noisy Fock states, but does not yield any advantage for the two other examples. This is due to the eigenvector drift phenomenon (see Section~\ref{sec:background_VD}).}
    \label{fig:number_op}
\end{figure*}

In order to test the effectiveness of the virtual distillation protocols introduced in the last section, we consider different sources of noise that would affect the ideal states in a practical setup. We will represent the noisy state as the output of a quantum channel acting on the ideal state $|\psi_{0}\rangle$ as 
\begin{equation}\label{eq:rho channel}
    \rho=\mathcal{E}\left(|\psi_{0}\rangle\langle \psi_{0}|\right)=\sum_{j}\hat{E}_{j}|\psi_{0}\rangle\langle \psi_{0}|\hat{E}_{j}^{\dagger},
\end{equation}
where $\{\hat{E}_{j}\}$ is the set of Kraus operators representing the quantum channel. Two of the main sources of noise in CV systems come from loss and dephasing errors \cite{Joshi_2021}, which are defined in terms of Kraus operators as
\begin{align*}
&\hat{A}_k(t)=\dfrac{(1-e^{-\kappa t})^{k/2}}{\sqrt{k!}}\;e^{-\kappa \hat n t/2}\;\hat a^k \quad & \text{(Loss)}\\
&\hat B_\ell(t) = \frac{(\gamma t)^{\ell/2}}{\sqrt{\ell!}}e^{-\gamma\hat n^2t/2}\hat n^\ell \quad & \text{(Dephasing)}
\end{align*}
for all $k,\ell\in \mathbb{N}$. Here $\kappa$ and $\gamma$ are  the loss and pure dephasing rates respectively.

Note that after evolving the target state through a channel like in Eq.~\eqref{eq:rho channel}, the dominant eigenvector of $\rho$ may not be $\ket{\psi_0}$ anymore. This scenario, in which the starting assumption Eq.~\eqref{eq:noisy-input} is not fulfilled, is precisely the eigenvector drift phenomenon mentioned in Sec.~\ref{sec:background_VD}, and we will further comment on it throughout the examples.

This Section is devoted to testing the VD protocols introduced in the previous section for various combinations of target observable, input state, and noise type. In Appendix~\ref{sec:noise-resilience}, we further assess the robustness of the protocols to additional errors that may occur in the quantum circuit implementing VD itself, both from systematic coherent errors in the gates and from incoherent losses and dephasing occurring in the modes. A similar analysis of the robustness of VD for qubits has been performed in Ref.~\cite{vikstaal2024study}.

\subsection{Photon number measurements under photon losses}\label{sec:number_op_lossy_Fock}

We start by considering the case of the number operator, whose expectation value we wish to estimate  on some specific states, using the first method for VD. We first look at input Fock states in the presence of pure losses, since dephasing does not affect these specific states. Fig.~\ref{fig:number_op}(a) summarizes the result of the virtual distillation protocol on Fock state $\ket3$ under increasing loss rate. We observe a clear advantage in applying virtual distillation, meaning that expectation values obtained using the protocol are significantly closer to the ideal value as compared to the case without virtual distillation. Moreover, we see that increasing the number of copies $M$ leads to improved results, as expected from our analysis. We refer to Appendix~\ref{sec: example} for a detailed analysis of the very similar case of Fock state $\ket1$.

However, when looking at other states that are far from eigenstates of $\hat{n}$, we observe that the protocol does not yield any advantage. This is illustrated in Fig.~\ref{fig:number_op}(b) and Fig.~\ref{fig:number_op}(c), respectively for a binomial state with $\ket{\psi}_b = \frac{1}{\sqrt{2}}(\ket{0} + \ket{4})$, and a cat state with $\ket{\psi}_c \propto \ket{\alpha} + \ket{-\alpha}$ (with $\alpha = 1$). This behavior is a consequence of the eigenvector drift phenomenon: under a loss channel, a coherent state with amplitude $\ket{\alpha}$ evolves as $\hat\rho(t)=\ketbra{\alpha e^{-\kappa t/2}}$ making the dominant vector $\ketbra{\alpha}$ drift as a function of time. Similarly, for the binomial state, the main eigenvector shifts from $\ket{\psi_b}$ to  $\ket{\psi_b(t)}\propto\ket0+e^{-2\kappa t}\ket4$, and the distillation protocol distills with respect to the new eigenvector instead of the target state. Effectively, this boils down to the fact that the input states are not of the form of Eq.~\eqref{eq:noisy-input}.
When $\ket{\psi_0}$ is an eigenstate of $\hat{O}$, the drift of the eigenvector is less important, as compared e.g.\ to  incoherent amplitude loss, which is consistent with the findings of Ref.~\cite{Koczor_2021}. This intuitively explains why the protocol for estimating the expectation value of $\hat n$ performs at its best on Fock states, which are eigenstates of the target observable $\hat n$.

\subsection{Parity measurements and phase shift operator under photon losses}

\begin{figure*}[!tbhp]
    \centering
    \includegraphics[width=1\linewidth]{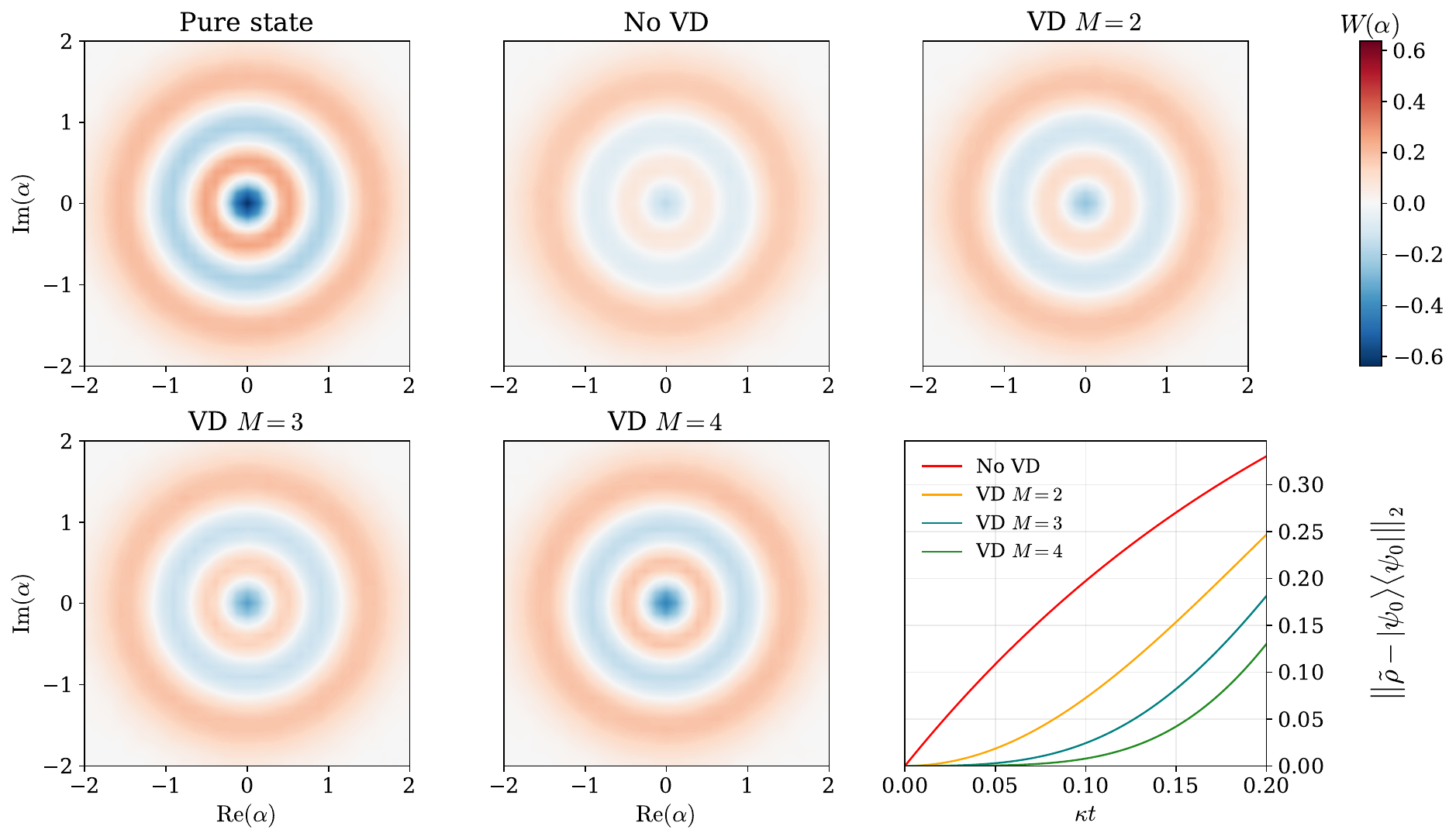}
    \caption{Reconstruction of the Wigner function $W(\alpha)$ for the Fock state $|3\rangle$ under a loss channel with strength $\kappa t= 0.2$. The heatmaps display the Wigner functions for the unmitigated noisy state (with label "No VD"), and the mitigated states using virtual distillation (labeled as "VD") with $M=2$, $3$, and $4$ copies. The measurements are obtained via displaced parity measurements. The right panel shows the phase-space $L_2$ error (Euclidean distance, see Eq.\eqref{eq: Plancherel}) relative to the ideal pure state's Wigner function as a function of the loss time $t$. As shown, increasing the number of copies $M$ effectively reduces the distance to the ideal state, demonstrating an improvement as compared to the case where no VD error mitigation is applied.}
    \label{fig: Wigner 3}

    \centering
    \includegraphics[width=1\linewidth]{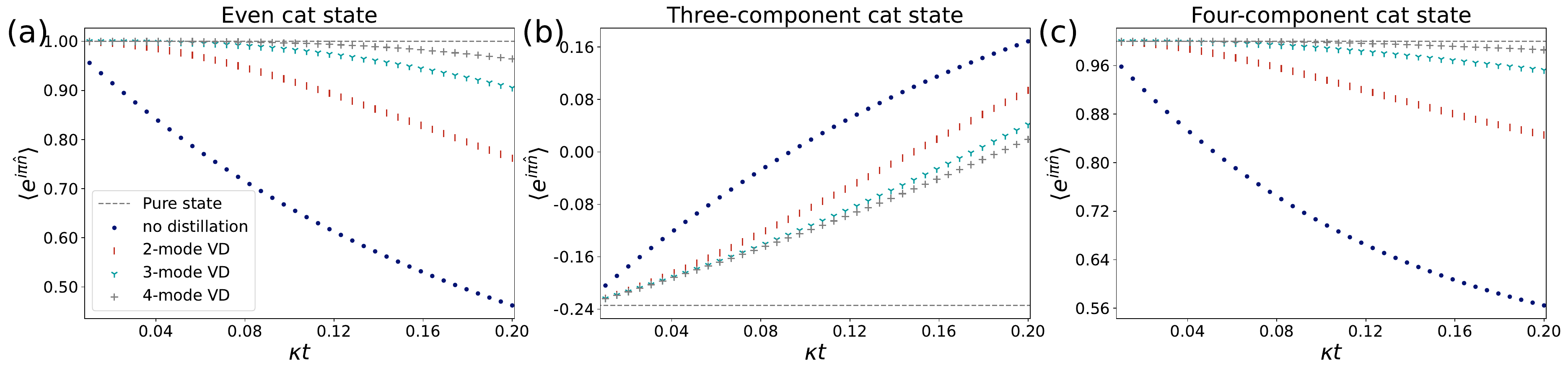} 
    \caption{
    Expectation value of the parity operator  $(-1)^{\hat{n}}$ as a function of the dimensionless loss rate for three noisy cat states with an amplitude of $\alpha=1.50$. The panels show the even cat ($n=2$), the three-component cat ($n=3$), and the four-component cat ($n=4$). The virtual distillation (VD) protocol performs best on the even and four-component cat states due to the fact that they are eigenstates of the target observable. For the three-component cat state, whose parity expectation value strongly depends on the value of $\alpha$, the protocol does not perform optimally but still achieves an improvement over the unmitigated noisy state.}
    \label{fig:Cats}
\end{figure*}

We now consider the estimation of the expectation value of the phase shift operator  $e^{i\phi\hat{n}}$, for which we rely on the second method for VD that we introduced. 
For the case of $\phi = \pi$, the phase shift operator $e^{i\phi\hat{n}}$ yields the parity operator $(-1)^{\hat n}$. The estimation of purified expectation value of the parity operator finds multiple applications.
One such application is obtaining the Wigner function
\begin{equation}
    W(\alpha)=\frac{2}{\pi}\Tr{\rho \hat{D}(\alpha)(-1)^{\hat{n}}\hat{D}(\alpha)^{\dagger}},
\end{equation}
where $\hat{D}(\alpha)$ is the displacement operator. If we are interested in obtaining the Wigner function of $|\psi_{0}\rangle$, we can use $M$ noisy copies $\rho$, displace them ($\hat{D}(\alpha)^{\dagger}\rho\hat{D}(\alpha)$) and use VD to approach the ideal Wigner function through parity measurements. In Fig.~\ref{fig: Wigner 3},  we show this procedure for the case of the state $|3\rangle\langle 3|$ under the effect of a loss channel. Although in reality the VD protocol does not distill a state, access to the Wigner function allows us to estimate how close the state $\Tilde{\rho}$ is to the ideal state we want to reconstruct. Indeed, by virtue of the quantum Plancherel identity \cite{becker2021convergence}, for 2 $m$-dimensional states $\rho_{1}$ and $\rho_{2}$, their Euclidean distance can be computed as
\begin{equation}\label{eq: Plancherel}
    \lVert \rho_{1}-\rho_{2}\rVert_{2}^{2}=\pi^{m}\int \left|W_{\rho_{1}}(\alpha)-W_{\rho_{2}}(\alpha)\right|^{2} d^{2m}\alpha.
\end{equation}
As one can see in Fig.~\ref{fig: Wigner 3}, increasing the number of copies reduces the distance from the ideal state, and not only the error on the observables. Estimating the parity operator $e^{i \hat n \pi}$ on displaced states $\hat D(\alpha) \rho \hat D^{\dagger}(\alpha)$ also finds application in the certification of stellar rank, by means of  stellar rank witnesses \cite{Oliver-paper}.

One can also consider the estimation of the expectation value of the operator  $e^{i\phi\hat{n}}$   on multi-component cat states, whose general form can be written as
\begin{equation}
    |\operatorname{cat}_{n}\rangle\propto\sum_{k=0}^{n-1}|e^{\frac{2\pi i}{n}k}\alpha\rangle
\end{equation}
for $\alpha\in \mathbb{C}$. Such states constitute the logical code-vector 0 of respective rotationally symmetric bosonic codes \cite{PhysRevX.10.011058}. In particular, in Fig.~\ref{fig:Cats}, we consider measurements of the parity operator for the cases $n=2,3,4$, which we refer to as (two-component) even cat, three-component cat and four-component cat state, evolved for different times under a loss channel. As one can see, the VD protocol performs the best on the even and four-component cat, due to the fact that they are eigenstates of the measured observable. The same does not hold for the three-component cat, whose parity does not have a definite value, and can only be defined in terms of its expectation value. Nevertheless, it is possible to notice that the protocol achieves an improvement for the parity, although not as significant as the other two cases.

\begin{figure*}[tbhp!]
    \includegraphics[width=1\linewidth]{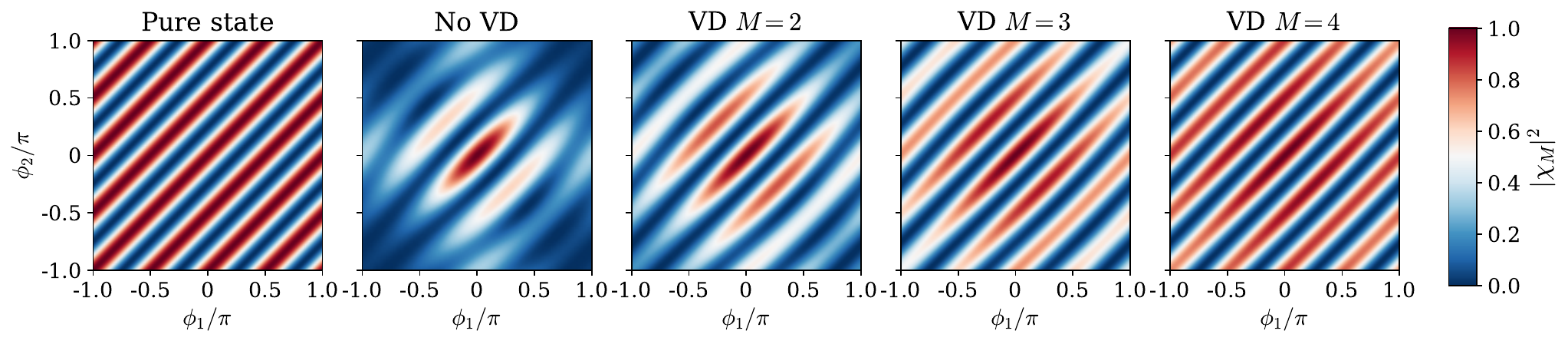}
    \caption{Modulus square of the characteristic function of the photon-number distribution of a two-mode $N00N$ state ($N=4$) after applying a loss channel with strength $\kappa t= 0.2$ .  Each point corresponds to a multimode estimation of the phases $e^{i(\phi_{1}\hat{n}_{1}+\phi_{2}\hat{n}_{2})}$. While the ideal properties of the characteristic function are quickly lost when noise is added, the virtual distillation (VD) protocol achieves an almost perfect reconstruction of the characteristic function describing the photon-number distribution on the two modes. This effective correction is possible because the $N00N$ states are eigenstates of the phase shifter operator.}
    \label{fig:N00N}

    \centering
    \includegraphics[width=\linewidth]{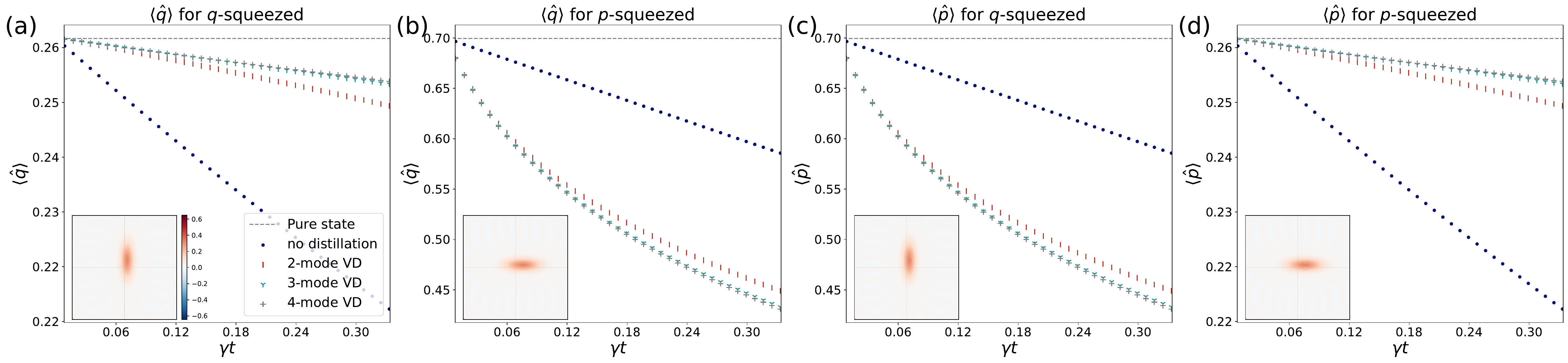} 
    \caption{
    The expected value of quadrature operators $\hat q$ ((a) and (b)) and $\hat p$ ((c) and (d)) as a function of the noise strength for input squeezed coherent states evolved under dephasing channel. The chosen parameters are $\alpha=0.3+0.3i$ and $\eta= 0.5$ in (a) and (c), and $\eta= -0.5$ in (b) and (d). The Wigner functions of the initial pure states are plotted accordingly. We observe that virtual distillation is valuable in the cases where the initial state is closer to an eigenstate of the measured observable (as it is the case in (a) and (d)), while it does not bring any advantage in the other case.}
    \label{fig:quadrature_op}
\end{figure*}

For two-component cat states~\cite{puri2017engineering}, the logical states 0 and 1 are encoded  in the even and odd photon-number subspaces, respectively. Estimating the parity therefore allows for accessing  information as to whether the cat qubit is in the zero or one logical state. Thus, by improving parity readout through virtual distillation, one can directly enhance the accuracy in the measurement of the Pauli $Z$ operator, with immediate application in the evaluation of the cost function used in variational quantum algorithms \cite{vikstal2024quantumapproximateoptimizationalgorithm}.

\subsection{Multimode phase shifts under photon losses}

Lastly, we show one example of multimode reconstruction of the photon-number characteristic function, focusing on the $N00N$ state, which has numerous applications in the context of quantum metrology \cite{Metrology_M_Barbieri}. Such states are defined as
\begin{equation}
    |N00N\rangle=\frac{|N0\rangle+|0N\rangle}{\sqrt{2}}
\end{equation}
where $N\in \mathbb{N}$. Due to the symmetries of this state, the photon-number characteristic function can be computed analytically as
\begin{align}
    \chi_{|N00N\rangle}(\phi_{1},\phi_{2})&=\langle N00N|e^{i(\phi_{1}\hat{n}_{1}+\phi_{2}\hat{n}_{2})}|N00N\rangle\\
    &=\frac{e^{iN\phi_{1}}+e^{iN\phi_{2}}}{2}
\end{align}
which yields $|\chi_{|N00N\rangle}(\phi,\phi)|^{2}=1$. Such a property is quickly lost under the effect of a loss channel, as we can see in Fig.~\ref{fig:N00N}. In particular, the addition of losses transforms the characteristic function in
\begin{align}
    \chi_{\text{loss}}(\phi_{1},\phi_{2})=\frac{1}{2}\sum_{j=1}^{2}\left[e^{-\kappa t + i\phi_{j}}+(1-e^{-\kappa t})\right]^{N}.
\end{align}

Nevertheless, since the $N00N$ states are eigenstates of the phase shifter operator, VD is able to correct for this, allowing for an almost perfect reconstruction of the photon-number characteristic function  (last panel of Fig.~\ref{fig:N00N}).

\subsection{Quadrature measurements under dephasing noise}

Finally, we consider as a last example  the quadrature operators $\hat q$ and $\hat p$, whose distilled expectation values we wish to retrieve using the first VD method we introduced. As input states, we choose to focus on squeezed coherent states, since they provide nontrivial expectation values for the quadrature operators --- whereas Fock states have zero expectation value for both quadratures.
Squeezed coherent states are generated by the squeezing operator 
\begin{equation}
    \hat{S}(\eta) = \exp\!\left(\tfrac{1}{2}\left(\eta^* \hat{a}^2 - \eta \hat{a}^{\dagger 2}\right)\right)
\end{equation}
applied on a coherent state $\ket\alpha$, where $\eta$ determines both the magnitude and direction of squeezing. For a phase $\arg(\eta)=0$, the squeezing occurs along the $\hat{q}$ quadrature, while for $\arg(\eta)=\pi$, it occurs along the $\hat{p}$ quadrature. 

However,  under photon  losses, such states are affected by strong eigenvector drift, making VD ineffective in this scenario. Therefore, we instead analyze the effectiveness of VD for a different kind of noise on the inputs, namely dephasing noise. The results are illustrated in Fig.~\ref{fig:quadrature_op}, where in panels (a) and (b), we evaluate the expectation value of the operator $\hat{q}$ for squeezed states with $\eta = 0.5$ and $\eta = -0.5$, respectively, and the same for $\hat p$ in panels (c) and (d).

Focusing on panel (a), the coherent state is squeezed along the $\hat{q}$-quadrature and therefore aligns with the measured quadrature. We again observe a significant advantage of using the virtual distillation protocol. In contrast, panel (c) shows that virtual distillation does not improve the measurement of $\hat p$, rather it gives worse results than a direct measurement. For a state squeezed in the $\hat p$ direction, the behavior is reversed: the state in panel (d), squeezed with $\eta = -0.5$ is closer to an eigenstate of $\hat{p}$, and thus yields better results. Once again, we observe the connection between the performance of the protocol and the closeness of the target state to an eigenstate of the measured observable.

\section{Conclusion}

In this work, we have applied the virtual distillation protocol in a bosonic framework providing a practical implementation of error mitigation in continuous-variable quantum systems. 
We provided two protocols for bosonic virtual distillation that can be implemented using passive linear interferometric networks. Our results show that expectation values for critical bosonic observables —including photon number, phase shift operators $e^{i \phi \hat n}$, as well as arbitrary quadratures of the bosonic field — can be successfully recovered from multi-copy measurements. We validated the effectiveness of these methods against realistic continuous-variable noise channels, such as photon loss and dephasing, highlighting substantial improvements for relevant states like Fock, squeezed, and cat states. In analogy with previous protocols for qubits \cite{Koczor_2021}, these improvements are most effective in the cases where the input state is an eigenstate of the operator whose expectation value is to be estimated. Furthermore, we outlined immediate practical applications for our protocol, notably in enhancing parity measurements for high-fidelity reconstruction of the Wigner function, and particle number correlators for quantum simulation.

Looking ahead, several open questions and future avenues of research remain. A primary challenge of these approaches is the impact of coherent eigenvector drift—where the dominant eigenvector of the noisy mixture drifts away from the target ideal state, invalidating the starting assumption Eq.~\eqref{eq:noisy-input}. This effect currently poses limits to the applicability of our methods, and warrants deeper theoretical and experimental exploration to extend its effectiveness to a broader class of  continuous-variable states and noise models.

Furthermore, while we have so far provided a protocol for the purified estimation of higher-order moments of the number operators and the photon-number characteristic function, it would also be desirable to provide an estimation protocol for  higher moments of the quadrature operators and the classical characteristic function $\langle e^{i r \hat q_\theta }\rangle $, for arbitrary quadratures $\hat q_\theta$ (see Eq.~\eqref{eq:arbitrary-quadrature}). Such a capability would find application in the error mitigated detection of non-Gaussianity~\cite{kala2025nullifiersnongaussianclusterstates}, as well as non-classicality \cite{PhysRevA.65.033830}.

\section*{Acknowledgments}
L.F. acknowledges funding from Chalmers Area of Advance Nano and Erasmus+ programs, resources at the Chalmers Centre for Computational Science and Engineering (C3SE), as well as funding from Institut Polytechnique de Paris (AMX). M.R. is a FRIA grantee of the Fonds de la Recherche Scientifique – FNRS. D. Gomes acknowledges funding by the Foundation of Science and Technology (FCT).  L.N., G. F. and D. Gunn acknowledge funding from the European Union’s Horizon Europe Framework Programme (EIC Pathfinder Challenge project Veriqub) under Grant Agreement No. 101114899. L.N.  acknowledges funding from FCT-Fundação para a Ciência e a Tecnologia (Portugal) via the Project No. CEECINST/00062/2018. L.N., D. Gomes and D. Gunn acknowledge support from the project with the reference n.º 2023.15565.PEX, funded by national funds through FCT – Fundação para a Ciência e a Tecnologia, I.P.
G.F. acknowledges financial support from the Swedish Research Council (Vetenskapsradet) through the project grant DAIQUIRI, as well as from the Olle Engkvist foundation. G.F. and A.U. acknowledge support from the Knut and Alice Wallenberg Foundation through the Wallenberg Center for Quantum Technology (WACQT).

\section*{Data Availability Statement}
Numerical data supporting the findings of this study are available at 
\href{https://github.com/lfinocchiaro/virtual-distillation}{\texttt{github.com/lfinocchiaro/virtual-distillation}}.

\bibliography{bibliography}

\appendix

\section{Sample complexity}\label{sec: Sample complexity}
In the above discussion, we did not consider the sample complexity of estimating $\Tr{\hat{O}\Tilde{\rho}}$, which of course will depend on the number of copies involved. Let us separate the numerator $\mathcal{N}=\Tr{\hat{O}_{1}\hat{S}^{(M)}\rho^{\otimes M}}$ and denominator $\mathcal{D}=\Tr{\hat{S}^{(M)}\rho^{\otimes M}}$. We start by considering the case in which numerator and denominator get estimated by two distinct measurements and thus independent, since it allows for better estimation bounds, later we will show the case in which the independence assumption is dropped. Suppose \(\mathcal N\) and \(\mathcal D\) are expectations of bounded random variables, estimated by sample means:
\begin{equation}
\hat{\mathcal N}=\frac{1}{n_N}\sum_{i=1}^{n_N} Y_i,\qquad\hat{\mathcal D}=\frac{1}{n_D}\sum_{j=1}^{n_D} Z_j,
\end{equation}
with
\begin{equation}
Y_i\in[a,b],\qquad Z_j\in [-1,1],
\end{equation}
where for $Z$ we measure observables of the type $e^{i\phi\hat{n}}$ which real part is bounded in $[-1,1]$, and the imaginary one can be neglected since $\mathcal{D}$ is real valued. Let us call $|\hat{\mathcal{N}}-\mathcal{N}|=\epsilon_{\mathcal{N}}$ and $|\hat{\mathcal{D}}-\mathcal{D}|=\epsilon_{\mathcal{D}}$, then
\begin{equation}
\left|\frac{\hat{\mathcal{N}}}{\hat{\mathcal{D}}}-\frac{\mathcal{N}}{\mathcal{D}}\right| \le\frac{\epsilon_{\mathcal{N}}}{\mathcal{D}-\epsilon_{\mathcal{D}}}+\frac{|\mathcal N|\,\epsilon_{\mathcal{D}}}{\mathcal{D}(\mathcal{D}-\epsilon_{\mathcal{D}})} + O\left(\epsilon_{\mathcal{N}}^2,\epsilon_{\mathcal{N}}\epsilon_{\mathcal{D}},\epsilon_{\mathcal{D}}^{2} \right).
\end{equation}
Truncated at first order in both $\epsilon_{\mathcal{N}}$ and $\epsilon_{\mathcal{D}}$. To obtain the bound
\begin{equation}\left|\frac{\hat{\mathcal{N}}}{\hat{\mathcal{D}}}-\frac{\mathcal{N}}{\mathcal{D}}\right|\le \varepsilon \end{equation}
we start by enforcing $2\epsilon_{\mathcal{D}}\leq \mathcal{D}$. As a consequence we can rewrite it as 
\begin{equation}
    \left|\frac{\hat{\mathcal{N}}}{\hat{\mathcal{D}}}-\frac{\mathcal{N}}{\mathcal{D}}\right|\leq \frac{2\epsilon_{\mathcal{N}}}{\mathcal{D}}+\frac{2|\mathcal{N}|\epsilon_{\mathcal{D}}}{\mathcal{D}^2} \leq \varepsilon \ .
\end{equation}
To simplify the condition above, we impose
\begin{align}
    \frac{2\epsilon_{\mathcal{N}}}{\mathcal{D}} \leq \alpha\varepsilon \implies &\epsilon_{\mathcal{N}} \leq \alpha\varepsilon\frac{\mathcal{D}}{2} \\
    \frac{2|\mathcal{N}|\epsilon_{\mathcal{D}}}{\mathcal{D}^2} \leq (1-\alpha)\varepsilon \implies &\epsilon_{\mathcal{D}} \leq (1-\alpha)\varepsilon\frac{\mathcal{D}^{2}}{2|\mathcal{N}|}  \ .
\end{align}
with $\alpha\in [0,1]$. If we apply Hoeffding's inequality that states that given $n$ i.i.d random variables $Z_{i}\in [a,b]$ almost surely with average $\mathbb{E}[Z_{i}]=\mu \ \forall i$ then 
\begin{equation}
    \operatorname{Pr}\left(\left|\frac{1}{n}\sum_{i}Z_{i} - \mu \right|\geq \varepsilon\right)\leq 2e^{-\frac{2n\varepsilon^{2}}{(b-a)^2}}   .
\end{equation}
Thus, to have probability $1-\delta$ to be $\varepsilon$ close to the value $\mu$ we can simply solve
\begin{equation}
    2e^{-\frac{2n\varepsilon^{2}}{(b-a)^2}} \leq \delta \implies n\geq \frac{(b-a)^2}{2\epsilon^2}\ln{\frac{2}{\delta}}.
\end{equation}
We can bound the number samples we need for both as
\begin{align}
    n_N &\ge \frac{2(b-a)^2}{\alpha^{2}\varepsilon^2\mathcal{D}^2}\ln\left(\frac{2}{\delta}\right) \\
    n_D &\ge  \frac{8|\mathcal{N}|^2}{(1-\alpha)^{2}\varepsilon^2\mathcal{D}^4}\ln\left(\frac{2}{\delta}\right) \ .
\end{align}
We can find the optimal value of $\alpha$ by minimizing $n_{N}+n_{D}$ as a function of $\alpha$. In particular, we define the function 
\begin{equation}
    f(\alpha)=\min (n_{N}+n_{D})= \frac{2\ln{\frac{2}{\delta}}}{\varepsilon^2}\left(\frac{(b-a)^2}{\alpha^2 \mathcal{D}^2}+\frac{4|\mathcal{N}|^2}{(1-\alpha)^2}\right)
\end{equation}
which, when we look for stationary point gives us
\begin{equation}
    \arg\left(\frac{d}{d\alpha}f(\alpha)=0\right)=\frac{1}{1+\left(\frac{2|\mathcal{N}|\mathcal{D}}{(b-a)} \right)^{\frac{2}{3}}} \ .
\end{equation}
The solution can be proved to be a minimum and a well defined solution since it lies in the interval $[0,1]$. So the total number of needed copies scale to be $\varepsilon$ with probability $1-\delta$ is bounded by
\begin{equation}
    n_{\operatorname{tot}}=n_{N}+n_{D}\geq \frac{2\ln{\frac{2}{\delta}}}{\varepsilon^{2}\mathcal{D}^{2}}\left[\left(\frac{b-a}{\mathcal{D}}\right)^{2/3} +\left(2|\mathcal{N}|\right)^{2/3} \right]^{3}
\end{equation}
which is strongly dependent on $\mathcal{D}^{-2}=\Tr{\rho^{M}}^{-2}\approx(1-\epsilon)^{-2M}$ . This implies that, even though the protocol allows for an exponential error suppression, this comes at an exponential cost in the number of copies. 

In the case where we estimate simultaneously both $\mathcal{N}$ and $\mathcal{D}$, i.e. Method~1, it is useful to compute the variance of their estimator ratio
\begin{equation}
    \operatorname{Var}\left(\frac{\hat{\mathcal{N}}}{\hat{\mathcal{D}}}\right)\approx \frac{1}{n}\left(\frac{\sigma_{N}^{2}}{\mathcal{N}^2}+ \frac{\sigma_{D}^{2}\mathcal{N}^{2}}{\mathcal{D}^{4}}-2\frac{\mathcal{N}}{\mathcal{D}^{3}}\operatorname{Cov}(\mathcal{N},\mathcal{D})\right)
\end{equation}
where $\mathbb{E}[\hat{\mathcal{D}}]=\mathcal{D},\mathbb{E}[\hat{\mathcal{N}}]=\mathcal{N}$, $\sigma_{D}$ and $\sigma_{N}$ are the respective standard deviations and $\operatorname{Cov}(A,B)$ is the covariance between $A$ and $B$. At this point, we can make use again of Hoeffding's inequality for sub-Gaussian variables and obtain that the number of samples needed to be $\varepsilon$ close with probability $1-\delta$ is:
\begin{equation}
     n_{\operatorname{tot}}\geq \frac{\frac{\sigma_{N}^{2}}{\mathcal{N}^2}+ \frac{\sigma_{D}^{2}\mathcal{N}^{2}}{\mathcal{D}^{4}}-2\frac{\mathcal{N}}{\mathcal{D}^{3}}\operatorname{Cov}(\mathcal{N},\mathcal{D})}{\varepsilon^2}\ln{\frac{1}{\delta}}   .
\end{equation}
The scaling is exponential as before, but in this case we notice that there is a dependence on the covariance between numerator and denominator. In this case though, it is not possible to quantify the number of samples needed for $\mathcal{N}$ and the one for $\mathcal{D}$. 

While preparing this manuscript, we noted a very recent analysis of sample complexity by V. Scavino Alfaro \cite{alfaro2026certified}, which explores this topic within the specific context of qubit systems.

\section{Illustrative Example: Single-Photon State Under Loss}\label{sec: example}
\begin{figure}[thbp!]
    \centering
    \includegraphics[width=1\linewidth]{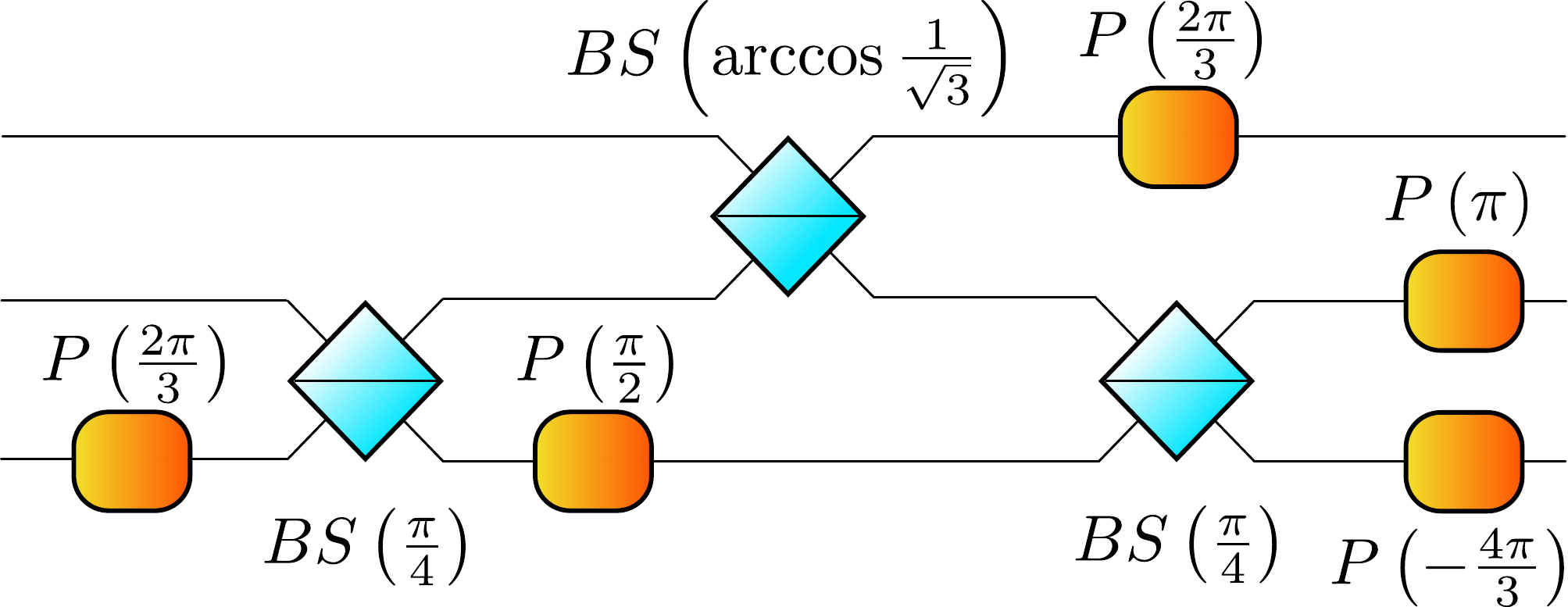}
    \caption{
Circuit decomposition \cite{hamiltonian_theory_multiport,multiport_bs} of the 3-mode Fourier transform \(\mathcal{F}_3\) into three beam splitters and local phase shifts. Here \(P(\phi)\) denotes a single-mode phase shifter applying a phase \(\phi\), i.e. $e^{i\phi\hat{n}_{j}}$, and \(BS(\theta)\) denotes a beam splitter with transmissivity \(T=\cos^2\theta\) under the standard convention. 
}
    \label{fig: Fourier_3}
\end{figure}

To give an intuition of the protocol we work out explicitly one example, which can be directly computed. We consider the case of a state $|\psi_{0}\rangle=|1\rangle$ under the effect of a loss channel, such that the noisy state can be written as 
\begin{equation}
    \rho=(1-\epsilon)|1\rangle\langle 1|+\epsilon|0\rangle\langle 0|
\end{equation}
with $\epsilon> \frac{1}{2}$ to that the dominant eigenvector is the target state. We consider the case of $M=3$ copies. We can then write the Fourier matrix as 
\begin{equation}
    \mathcal{F}_{3}=\frac1{\sqrt3}\begin{bmatrix} e^{i\frac{2\pi}{3}} & e^{-i\frac{2\pi}{3}} &1 \\ e^{-i\frac{2\pi}{3}} & e^{i\frac{2\pi}{3}} &1 \\ 1&1&1\end{bmatrix}
\end{equation}
and the eigenvalues of $\hat S^{(3)}$ are $\{1,e^{i\frac{2\pi}{3}},e^{-i\frac{2\pi}{3}}\}$. This gate can be implemented as a linear interferometer given in Fig.~\ref{fig: Fourier_3}. 

Since we have a mixture of single photons and vacuum, we can explicitly write down the probabilities in terms of matrix permanents. In particular, as shown in Ref.~\cite{Tichy_2014}, the transition probability from an input occupation vector $\boldsymbol{m}=(m_{1},...,m_{k})$ (meaning $m_{1}$ bosons in mode $1$, $m_{2}$ bosons in mode $2$, etc.) to an output configuration vector $\boldsymbol{k}$ can be expressed in terms of matrix permanents as
\begin{equation}
    p_{\boldsymbol{m}\to \boldsymbol{k}}=\frac{|\operatorname{perm}(U_{\boldsymbol{m},\boldsymbol{k}})|^{2}}{\boldsymbol{k}!\boldsymbol{m}!},
\end{equation}
where $U_{\boldsymbol{m},\boldsymbol{k}}$ is the relevant scattering matrix of $U$ obtained by repetition of rows and column according to the multiplicity of $\boldsymbol{m},\boldsymbol{k}$ and the permanent of a matrix is defined as
\begin{equation}
    \operatorname{perm}(A)=\sum_{\sigma \in S_{n}}\prod_{i=1}^{n}A_{i,\sigma_{i}}
\end{equation}
with $S_{n}$ the symmetric group of $n$ elements. So the case of $\rho^{\otimes 3}$ as an input, the probability of measuring the output $\boldsymbol{k}$ can be written as
\begin{align}
    P(\boldsymbol{k})&=\sum_{\boldsymbol{m}\in \{0,1\}^{3}} p_{\boldsymbol{m}\to \boldsymbol{k}}\prod_{i=1}^{3}\epsilon^{1-m_{i}}(1-\epsilon)^{m_{i}} \\
    &= \sum_{\boldsymbol{m}\in \{0,1\}^{3}} \frac{|\operatorname{perm}(U_{\boldsymbol{m},\boldsymbol{k}})|^{2}}{\boldsymbol{k}!}\prod_{i=1}^{3}\epsilon^{1-m_{i}}(1-\epsilon)^{m_{i}}   .
\end{align}
We show the result in the following table: 
\[
\begin{array}{|c|c|}
\hline
 \bm{k}=(k_1,k_2,k_3) &  P(\bm{k}) \\[1pt] \hline 
  & \\[1 pt]
(0,0,0)  \quad& \quad\epsilon^3 \\[4pt]
(1,0,0),\ (0,1,0),\ (0,0,1) \quad & \quad \epsilon^2(1-\epsilon) \\[4pt]
(2,0,0),\ (0,2,0),\ (0,0,2) \quad & \quad\dfrac{2}{3}\epsilon(1-\epsilon)^2 \\[8pt]
(1,1,0),\ (1,0,1),\ (0,1,1) \quad & \quad\dfrac{1}{3}\epsilon(1-\epsilon)^2 \\[8pt]
(1,1,1) \quad  & \quad\dfrac{1}{3}(1-\epsilon)^3 \\[8pt]
(3,0,0),\ (0,3,0),\ (0,0,3) \quad& \quad\dfrac{2}{9}(1-\epsilon)^3 \\[8pt]
(2,1,0),\ (2,0,1) ,\ (0,1,2) \quad& \quad 0 \\[8pt]
(1,2,0),\ (1,0,2) ,\ (0,2,1) \quad& \quad 0 \\[8pt]
\hline
\end{array}
\]
To compute $\Tr{\hat{n}^{(3)}\hat{S}^{(3)}\rho^{\otimes 3}}$ and $\Tr{\hat{S}^{(3)}\rho^{\otimes 3}}$, we can simply compute
\begin{align}
    \Tr{\hat{S}^{(3)}\rho^{\otimes 3}}&=\sum_{\bm{k}}P(\bm{k}) e^{\frac{2\pi i}{3}\sum_{j=1}^{3}jk_{j}} \\
     &= \epsilon^{3}+(1-\epsilon)^{3} \quad ,\\
    \Tr{\hat{n}^{(3)}\hat{S}^{(3)}\rho^{\otimes 3}}&=\frac{1}{3}\sum_{\bm{k}}P(\bm{k})|\bm{k}| e^{\frac{2\pi i}{3}\sum_{j=1}^{3}jk_{j}}\\
    &=(1-\epsilon)^{3} 
\end{align}
with $|\bm{k}|=\sum_{j}k_{j}$. By combining the two we obtain
\begin{align}
    \frac{\Tr{\hat{n}^{(3)}\hat{S}^{(3)}\rho^{\otimes 3}}}{\Tr{\hat{S}^{(3)}\rho^{\otimes 3}}}&=\frac{(1-\epsilon)^{3}}{\epsilon^{3}+(1-\epsilon)^{3}}\\
    &=\frac{1}{1+\left(\frac{\epsilon}{1-\epsilon}\right)^{3}} \approx 1- O(\epsilon^{3}) 
\end{align}
which, coherently with the above discussion, is $\langle \psi_{0}|\hat{n}|\psi_{0}\rangle=\langle 1|\hat{n}|1\rangle=1$ with an error of order $\epsilon^{3}$.

We can repeat the same calculation for the parity. Here, we have that
\begin{equation}
        (-1)^{\hat n_1}\hat{S}^{(M)} =\begin{bmatrix} 
0 & 0 & -1 \\ 
1 & 0 & 0 \\ 
0 & 1 & 0
\end{bmatrix}.
\end{equation}
When we diagonalize it we obtain
\begin{align}
    D = \begin{pmatrix} e^{i\pi} & 0 & 0 \\ 0 & e^{-i\frac{\pi}{3}} & 0 \\ 0 & 0 & e^{i\frac{\pi}{3}} \end{pmatrix}\\
V = \frac{1}{\sqrt{3}} \begin{pmatrix} 1 & e^{-i\pi} & 1 \\ e^{i\frac{2\pi}{3}} & e^{i\frac{\pi}{3}} & 1 \\ e^{-i\frac{2\pi}{3}} & e^{-i\frac{\pi}{3}} & 1 \end{pmatrix}
\end{align}
such that $V^{\dagger}D V=(-1)^{\hat n_1}\hat{S}^{(M)}$. The matrix $V$ is not a Fourier transform, but for the specific input state $\rho^{\otimes 3}$, it  gives rise to the same probabilities. Thus we can repeat the same type of calculation of before, and we obtain
\begin{align}
\Tr{e^{i\pi\hat{n}_{1}}\hat{S}^{(3)}\rho^{\otimes 3}}&=\frac{1}{3}\sum_{\bm{k}}P(\bm{k}) e^{i\pi \left(k_{1}-\frac{k_{2}-k_{3}}{3}\right)}\\
    &=-(1-\epsilon)^{3}.
\end{align}
By combining the two we obtain
\begin{align} \frac{\Tr{e^{i\pi\hat{n}_{1}}\hat{S}^{(3)}\rho^{\otimes 3}}}{\Tr{\hat{S}^{(3)}\rho^{\otimes 3}}}&=\frac{-(1-\epsilon)^{3}}{\epsilon^{3}+(1-\epsilon)^{3}}\\
    &=\frac{-1}{1+\left(\frac{\epsilon}{1-\epsilon}\right)^{3}} \approx -1+ O(\epsilon^{3}) 
\end{align}
which, coherently with the above theory, is $\langle \psi_{0}|e^{i\pi\hat{n}}|\psi_{0}\rangle=\langle 1|e^{i\pi\hat{n}}|1\rangle=-1$ with an error of order $\epsilon^{3}$.

\section{Bound on observable distance}\label{sec: Bound}
In this Appendix, we prove the bound provided in Eq.~\eqref{eq: bound via maxi} as well as an alternative bound that uses fewer assumptions, Eq.~\eqref{eq:error_bound_noAssump}.
Let us begin by restating Eq.~\eqref{eq: bound via maxi} for convenience: 
\begin{equation}
    \left|\Tr{\hat{O}\left(\tilde{\rho} - |\psi_{0}\rangle\langle \psi_{0}|\right)}\right|
    \leq 2\max_{i}\left|\Tr{\hat{O}|\psi_{i}\rangle\langle\psi_{i}|}\right|\,\mathcal{O}(\epsilon^{M}).
\end{equation}
\begin{proof}
    Given Eq.~\eqref{eq:noisy-input}, we can rewrite the error as:
    \begin{align}
        \left|\langle \hat{O}\rangle_{\psi_{0}}-\langle\hat{O}\rangle_{\tilde{\rho}}\right|&\leq \frac{\sum_{i}\epsilon_{i}^{M}|\langle \hat{O}\rangle_{\psi_{i}}|+\left(\sum_{i}\epsilon_{i}^{M}\right)|\langle \hat{O}\rangle_{\psi_{0}}|}{(1-\epsilon)^{M}+\sum_{i}\epsilon_{i}^{M}}\\
        &\leq \max_{i}\left|\langle \psi_{i}|\hat{O}|\psi_{i}\rangle\right| \frac{2\sum_{i}\epsilon_{i}^{M}}{(1-\epsilon)^{M}+\sum_{i}\epsilon_{i}^{M}}
    \end{align}
We then notice that given the assumption of $1-\epsilon\geq \max_{i}\epsilon_{i}$ we have
    \begin{align}\label{eq: epsilon inequality}
        \frac{\sum_{i}\epsilon_{i}^{M}}{(1-\epsilon)^{M}+\sum_{i}\epsilon_{i}^{M}} \leq \frac{\epsilon^{M}}{(1-\epsilon)^{M}+\epsilon^{M}}.
    \end{align}
    This can be easily shown by realizing that $\sum_{i}\epsilon_{i}^{M}\leq \epsilon^{M}$, and rewriting the above as
    \begin{equation}
        1-\frac{(1-\epsilon)^{M}}{(1-\epsilon)^{M}+\sum_{i}\epsilon_{i}^{M}}\leq 1-\frac{(1-\epsilon)^{M}}{(1-\epsilon)^{M}+\epsilon^{M}}.
    \end{equation}
    This inequality is proven by realizing that the function 
    \begin{equation}
        f(x)=1-\frac{(1-\epsilon)^{M}}{(1-\epsilon)^{M}+x}.
    \end{equation}
    is monotonically increasing for all $x\geq 0$. This is sufficient to prove Eq.~\eqref{eq: epsilon inequality}. So we can rewrite the above as
    \begin{align}
        \left|\langle \hat{O}\rangle_{\psi_{0}}-\langle\hat{O}\rangle_{\tilde{\rho}}\right|\leq 2\max_{i}\left|\langle \psi_{i}|\hat{O}|\psi_{i}\rangle\right| \mathcal{O}(\epsilon^{M})
    \end{align}
    which concludes the proof.

\end{proof}

If we want to make the fewest amount of assumption on the estimation error, we can use Eq.~\eqref{eq:noisy-input} to obtain the following bound:
\begin{equation}\label{eq:error_bound_noAssump}
    \left|\Tr{\hat{O}\left(\tilde{\rho} - |\psi_{0}\rangle\langle \psi_{0}|\right)}\right|
    \leq \sqrt{\langle \hat{O}^{2}\rangle_{\tilde{\rho}} + \langle \hat{O}^{2}\rangle_{\psi_{0}}}\,
    \mathcal{O}\!\left(\epsilon^{M/2}\right).
\end{equation}
\begin{proof}

Consider two states $\rho$ and $\sigma$ diagonal in the same basis, then we have that
\begin{equation}
    \rho=\sum_{k}p_{k}|\psi_{k}\rangle\langle \psi_{k}| \ \ , \ \  \sigma=\sum_{k}q_{k}|\psi_{k}\rangle\langle \psi_{k}|.
\end{equation}
The distance between two observables can be written then as
\begin{align}
    \Tr{\hat{O}(\rho-\sigma)}&=\sum_{k}\langle \psi_{k}|\hat{O}|\psi_{k}\rangle(p_{k}-q_{k})\\
    &=\sum_{k}o_{k}(p_{k}-q_{k})   .
\end{align}

We can bound such quantity by using Cauchy–Schwarz inequality
\begin{align}
    \left|\Tr{\hat{O}(\rho-\sigma)}\right|&\leq \sum_{k}|o_{k}||p_{k}-q_{k}|\\
    &= \sum_{k}\sqrt{o_{k}^{2}}\sqrt{(p_{k}-q_{k})^{2}}\\
    &= \sum_{k}\sqrt{o_{k}^{2}(p_{k}+q_{k})}\sqrt{\frac{(p_{k}-q_{k})^{2}}{p_{k}+q_{k}}}\\
    &\leq \sqrt{\sum_{k}o_{k}^{2}(p_{k}+q_{k})}\sqrt{\sum_{k}\frac{(p_{k}-q_{k})^{2}}{p_{k}+q_{k}}}  .
\end{align}
We notice that 
\begin{equation}
    (p_{k}-q_{k})^{2}\leq |p_{k}-q_{k}|(p_{k}+q_{k})
\end{equation}
which implies
\begin{equation}
    \sum_{k}\frac{(p_{k}-q_{k})^{2}}{p_{k}+q_{k}}\leq \sum_{k}|p_{k}-q_{k}|=\lVert\rho-\sigma  \rVert_{1}   .
\end{equation}

As a consequence we have that
\begin{align}
    \left|\Tr{\hat{O}(\rho-\sigma)}\right|&\leq \sqrt{\sum_{k}o_{k}^{2}(p_{k}+q_{k})}\sqrt{\lVert\rho-\sigma  \rVert_{1}}\\
    &\leq \sqrt{\Tr{\hat{O}^{2}\rho}+\Tr{\hat{O}^{2}\sigma}}\sqrt{\lVert\rho-\sigma  \rVert_{1}}   .
\end{align}

In the case we discuss in the main text, we have two states $|\psi_{0}\rangle$ and $\tilde{\rho}$, which have trace distance $\mathcal{O}\left(\epsilon^{M}\right)$, which leads to Eq.~\eqref{eq:error_bound_noAssump}. 
\end{proof}

\section{Analytical form of $\hat{V}_{\phi}$ and $\hat{D}_{\phi}$}\label{sec: Analytical form}
To find the analytical form of the matrix $\hat{V}_{\phi}$ and $\hat{D}_{\phi}$, we start by considering the matrix 
\begin{equation} 
\operatorname{diag}\left(e^{i\phi},1,\dots,1\right)S_{M}=
    \begin{bmatrix} 
0 & 0 & \dots & e^{i\phi} \\ 
1 & 0& \dots & 0 \\ 
0 & 1 & \dots & 0\\
0& \dots & 1 & 0
\end{bmatrix}.
\end{equation}
First we compute the characteristic polynomial, which reduces to 
\begin{equation}
    p(\lambda)=\lambda^{M}-e^{i\phi} \implies \lambda_{k}=e^{i\frac{(2\pi k + \phi )}{M}}
\end{equation}
meaning $D_{\phi}=\operatorname{diag}(\lambda_{1},\dots,\lambda_{M})$. The corresponding eigenvector of $\lambda_k$ can be found in the form 
\begin{equation}
    v_{k}=\frac{1}{\sqrt{M}}\begin{bmatrix}
        1 \\ \lambda_{k}^{-1} \\ \lambda_{k}^{-2}\\ \vdots \\ \lambda_{k}^{-M}
    \end{bmatrix}
\end{equation}
and as a consequence the matrix $V_{\phi}$ can be written as 
\begin{equation}
   ( V_{\phi})_{jk}=\frac{1}{\sqrt{M}}e^{-\frac{i(j-1)}{M}(2\pi k +\phi)} \ .
\end{equation}
As expected, the matrix $V_0$ is a DFT in the case $\phi=0$.

\section{Moments estimation from the characteristic function}\label{sec: Moment estimation}

As mentioned above, it is possible to estimate the moment of the photon  number distribution from the characteristic function. In particular, for an $m$-mode state, we have two possible approaches: (i) we can take derivatives of the characteristic function or (ii) we can reconstruct the full probability distribution for the modes we are interested in. 

We consider the derivative approach first, since it requires to evaluate the characteristic function in a smaller number of points than the $N^{k}$ to reconstruct the correlator of order $k$.
By definition we have that
\begin{equation}
    \frac{\partial^{|\boldsymbol{k}|}}{\partial \phi_{1}^{k_{1}}\dots \partial \phi_{m}^{k_{m}} }\chi(\phi_{1},...,\phi_{m}){\bigg |}_{\boldsymbol{\phi}=0}=\operatorname{i}^{|\boldsymbol{k}|}\langle \hat{n}_{1}^{k_{1}}\dots\hat{n}_{m}^{k_{m}}\rangle
\end{equation}
with $\boldsymbol{k}=(k_{1},\dots,k_{m})\in \mathbb{N}^{m}$. We can use finite difference to compute the derivatives, but this comes with an error due to two different factors: we can only approximate the characteristic function up to additive error and the finite difference comes with an intrinsic error due to the truncation. Let us consider the case in which we want to measure the correlator $\langle \hat{n}_{i}\hat{n}_{j}\rangle$, where we will fix $(i,j)=(1,2)$ for simplicity. It can then be written as
\begin{align}
    \langle\hat{n}_{1}\hat{n}_{2}\rangle&=-\frac{\chi(\Delta \phi,\Delta \phi,0,\dots)+\chi(-\Delta \phi,-\Delta \phi,0,\dots)}{4(\Delta \phi)^{2}} +\\
    &+\frac{\chi(\Delta \phi,-\Delta \phi,0,\dots)+\chi(-\Delta \phi,\Delta \phi,0,\dots)}{4(\Delta \phi)^{2}}   .
\end{align}

The error associated with the truncation of the finite difference ($E_{\operatorname{tr}}$) can be computed as 
\begin{equation}
    E_{\operatorname{tr}}\leq \frac{1}{6}(\Delta \phi)^{2}\left(\langle \hat{n}_{1}^{3}\hat{n}_{2}+\hat{n}_{2}^{3}\hat{n}_{1}\rangle  \right) +O\left((\Delta \phi)^{3}\right)   .
\end{equation}
The error associated with the estimation of the characteristic function ($E_{\operatorname{est}}$) can be bound as
\begin{equation}
    E_{\operatorname{est}}\leq \frac{|\pm 4\varepsilon|}{4(\Delta \phi)^{2}}= \frac{\varepsilon}{(\Delta \phi)^{2}}   .
\end{equation}
Assuming the two errors are independent of each other, the total estimation error in the correlator ($E_{\operatorname{corr}}$) can be written as
\begin{align}
    E_{\operatorname{corr}}&\leq \frac{\varepsilon}{(\Delta \phi)^{2}} + \frac{1}{6}(\Delta \phi)^{2}\left(\langle \hat{n}_{1}^{3}\hat{n}_{2}+\hat{n}_{2}^{3}\hat{n}_{1} \rangle \right) \\
    &\leq \frac{\varepsilon}{(\Delta \phi)^{2}} + \frac{1}{3}(\Delta \phi)^{2}N^{4}
\end{align}
where we use the bound on the total number of particles to avoid the dependence on higher-order correlators. The optimal choice of $\Delta \phi$ can be found by minimizing the right hand side of the above, which gives us
\begin{equation}
    \Delta \phi_{opt}=\frac{\left(3\varepsilon\right )^{\frac{1}{4}}}{N}   .
\end{equation}
This leads to the bound
\begin{equation}
    E_{\operatorname{corr}} \leq 2N^{2}\sqrt{\frac{\varepsilon}{3}}   .
\end{equation}
With this approach, we only need to estimate $4+1$ quantities to have access to the second order correlators ($1$ for the $\Tr{\rho^{M}}$ and $4$ for the characteristic function), instead of the total $N^{2}$ that would be needed to reconstruct the full photon number probability distribution over two modes, from which the two-mode photon number correlator can be inferred. 
In general the same can be done for the order $k$-th order correlator, where $2^{k}+1$ measurements are needed. Additionally when we consider higher-order correlators, the dependence on $\varepsilon$ will be more significant.

Using the second method for estimating $\langle \hat{n}_{1}\hat{n}_{2}\rangle$ requires us to estimate the characteristic function in $N^{2}$ points, namely all tuples $\frac{2\pi}{N}(k_{1},k_{2}) \ \forall k_{1},k_{2}\in \{0,...,N-1\}$. If we consider the additive error on the estimation of the characteristic function, we get that
\begin{align}
    E_{\operatorname{corr}}&=  \left| \sum_{l_{1},l_{2}=0}^{N-1}\Delta\left(p(l_{1},l_{2})\right)l_{1}l_{2}\right|\\
    &\leq  \sum_{l_{1},l_{2}=0}^{N-1}\left|\Delta\left(p(l_{1},l_{2})\right) \right|l_{1}l_{2}\\
    &\leq  \varepsilon N^{2}
\end{align}
where we used the fact that an error of order $\varepsilon$ on the characteristic function translates into an error of order $\varepsilon$ on the probabilities. In general the error associated with the correlators of order $k$ will have a scaling of the type $O(N^{k}\varepsilon)$.

It is important to compare the two methods: using the derivative approach reduces the number of times we need to evaluate the characteristic function, which implies a smaller number of experimental measurements but it is associated with a higher error, for example in the case described above $\sqrt{\varepsilon}>\varepsilon$; on the other hand the inversion approach requires a large number of different measurements but the dependence of $\varepsilon$ is unchanged by the order of the correlators.

\begin{figure}[htbp!]
    {\centering
    \includegraphics[width=1\linewidth]{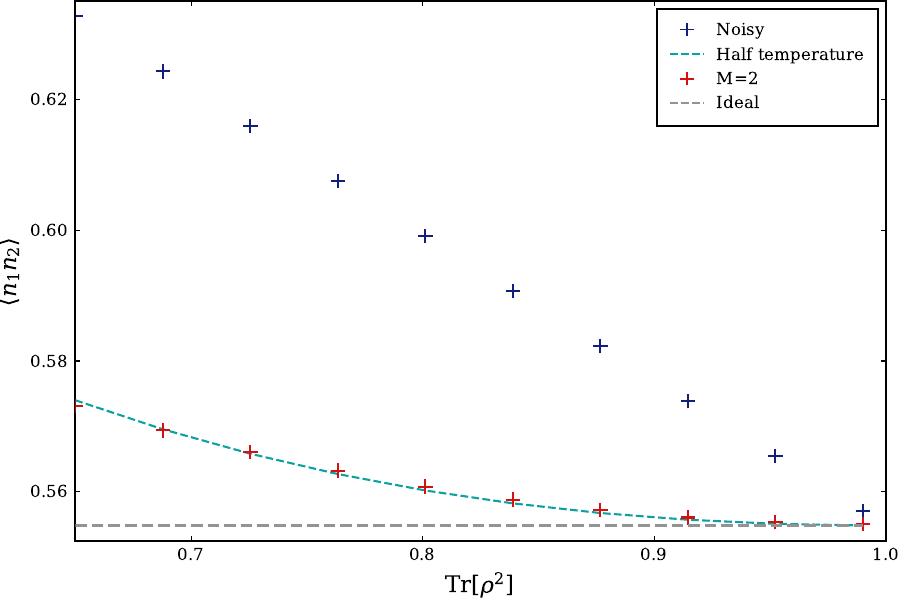}
    \centering\caption{Estimation of the particle number correlator $\langle \hat{n}_1 \hat{n}_2 \rangle$ as a function of the internal state purity $\Tr{\rho^2}=\tanh(\beta/2)$ for an Atomic Boson Sampling model. A random unitary is sampled from the Haar random average of size $4$ with $4$ particles. Bosonic particles evolve under a purely hopping Hamiltonian $\hat{H}=\sum_{ij}h_{ij}\hat{a}_{i}\hat{a}_{j}^{\dagger}$ in an optical lattice, where quantum interference is degraded by the phenomenon of partial distinguishability. To model this noise, the internal degrees of freedom of the particles are approximated as thermal states governed by an effective inverse temperature $\beta$. The plot contrasts the expectation values of the unmitigated system (\emph{Noisy}) against error-mitigated estimations. The \emph{Half temperature} curve represents an intermediate cooling simulation where $\beta \rightarrow 2\beta$. The $M=2$ curve demonstrates the performance of the virtual distillation protocol with two noisy copies, showing substantial error suppression and precisely matching the one obtained by halving the temperature, as predicted (up to small noise due to the sampling error). }
    \label{fig:Correlator}}
\end{figure}

To show the effectiveness, we consider a model usually discussed in the context of Atomic Boson Sampling \cite{Young_2024}, where bosonic particles evolve under a purely hopping Hamiltonian $H=\sum_{ij}h_{ij}a_{i}^{\dagger}a_{j}$ in an optical lattice. This approach is mathematically equivalent to the usual Boson Sampling scenario \cite{aaronson2010computationalcomplexitylinearoptics}, where the discussion is carried on in terms of linear unitaries. It is known that for this computational task correlators, especially lower order, can be computed classically in an efficient way \cite{anguita2025experimentalvalidationbosonsampling,robbio2026complementaritybosonicfermionicmanybody}.  One major source of error in boson sampling is associated with the concept of partial distinguishability\cite{Tichy_2014} , which is the phenomenon for which differences in the internal degrees of freedom between the interfering  bosonic particles hinder the effect of quantum interference. In general, if we consider that each particles has an internal degree of freedom described by the state $\rho$, then the correlators can be written (for $i\neq j$) as
\begin{align}
    \langle \hat{n}_{i}\hat{n}_{j}\rangle &= \sum_{k \neq l} |U_{ik}|^2 |U_{jl}|^2 + \Tr{\rho^2} U_{ik}^* U_{il} U_{jl}^* U_{jk} .
\end{align}

In an optical lattice, each boson is trapped in a local harmonic oscillator at non-zero temperature, and so the state of the internal degrees of freedom (the energy levels of the oscillator) can be approximated as a thermal state \cite{geller2025measuringmultiparticleindistinguishabilitygeneralized}. Hence, in this scenario we describe 
\begin{equation}
    \rho=\frac{e^{-\beta \hat{n}_{\operatorname{int}}}}{\Tr{e^{-\beta \hat{n}_{\operatorname{int}}}}}
\end{equation}
where $\hat{n}_{\operatorname{int}}$ refers to the energy level of the internal degrees of freedom, not to be confused with $\hat{n}_{i}$ which describe the number of atoms in the $i$-th well. As a consequence we have that
\begin{equation}
    \Tr{\rho^{2}}=\tanh\left(\beta/2\right)   .
\end{equation}

\begin{figure*}[tbhp!]
\centering
  \includegraphics[width=\textwidth]{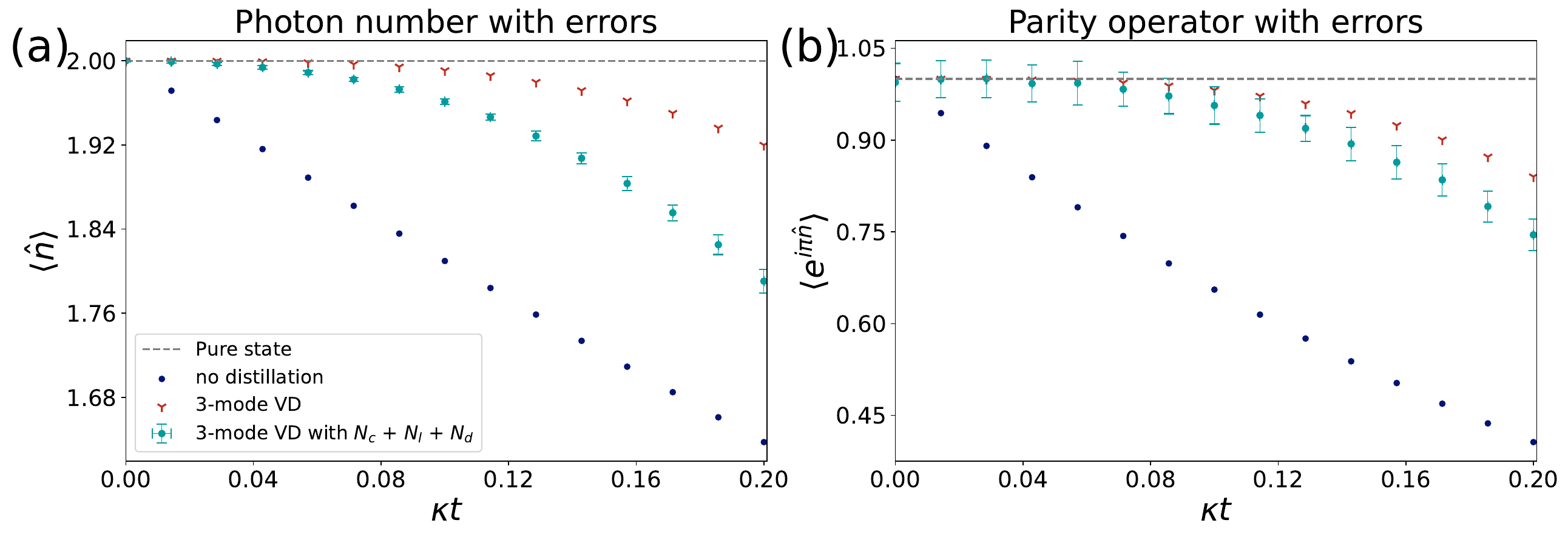}
    \caption{
    Expectation values of VD on input Fock states, with a noisy implementation of the protocol. We add both coherent error $N_c$ in the quantum gates (here with gaussian noise $\varepsilon\sim3\%$) and incoherent noise in each mode (losses $N_l$ and dephasing $N_d$), with strength $\sim 3\%$. The initial states are noisy Fock states $\ket2$ with losses, and in (a) we measure the number operator using Method~1 while in (b) we measure the parity operator using Method~2.
    }
    \label{fig:noisy_VD}
\end{figure*}

In Fig.~\ref{fig:Correlator}, we apply this characteristic-function approach to estimate second-order correlators in an Atomic Boson Sampling model, confirming that VD effectively counters partial distinguishability and approaches the ideal system behavior. This result is particularly compelling for two key reasons. First, it showcases the virtual distillation protocol's effectiveness against a fundamentally different and highly pertinent error model: partial distinguishability. While our earlier examples focused on standard continuous-variable noise channels such as amplitude loss and dephasing, here we demonstrate that the characteristic-function approach can successfully mitigate the degradation of quantum interference caused by mixed internal degrees of freedom. Second, although the present analysis is grounded in a non-interacting, purely hopping Hamiltonian where correlators can be evaluated classically, this method lays the groundwork for extensions to systems with atomic interactions. In interacting many-body regimes, where the analytical computation of exact observables quickly becomes intractable, our multi-copy estimation technique could serve as a critical experimental tool to reconstruct error-mitigated particle number correlators.

\section{Resilience to additional noise}\label{sec:noise-resilience}

In order to assess the robustness of the virtual distillation protocol, we consider the working scenario of number operator measurement on Fock states with losses using Method~1. We also consider parity operator measurement on the same input states using Method~2.

We perform a full simulation of the protocol using basic linear optics elements such as beam splitters and phase shifters, see e.g. Fig.~\ref{fig: Fourier_3}. We then perturb the protocol by adding two kinds of errors throughout the execution, namely \textit{incoherent} and \textit{coherent} errors. The first type refers to the effects of decoherence, which amounts to adding a loss and a dephasing channel independently to each mode and between each gate. The second type occurs when the parameters of the beam splitter gates $\operatorname{BS}(\theta)$ are slightly off due to an imprecise drive of the control inputs. To simulate the effects of such errors, we let the input states evolve through several samples of faulty implementations and perform statistics on the observed outcomes. More precisely, for each round, we draw random parameters $$\kappa_i,\gamma_i,\epsilon_i\sim\mathcal N,\quad i\in\{1,2,3\},$$ respectively for the decay and dephasing rate of mode $i$, and the error in the $i$-th beam-splitter. We consider normal distributions $\mathcal N$ with $3\%$ deviation from the average. Each mode evolves under noise channels with the chosen parameters between each gate, for a total time $\Delta t$ such that $\kappa\Delta t\sim\gamma\Delta t\sim3 \%$.

In Fig.~\ref{fig:noisy_VD}, we observe the expected results from this noisy setup. It appears that the protocol does not deviate too far from the theoretical value and, in particular, remains well above the result without distillation. Interestingly, the error bars appear larger in the second scenario where we measure the parity operator. This can be traced back to the fact that in Method~2, two parallel circuits must be executed, each with their own errors, while in Method~1, the final quantity is obtained by post-processing the data from a single circuit. It should be noted that the chosen error rates ($\sim3\%$) are much higher than the state of the art experimental capability \cite{Joshi_2021}, so we can conclude that the protocol is robust to additional coherent and incoherent errors.

\clearpage

\end{document}